\documentclass[fleqn,usenatbib]{mnras}

\usepackage{mathptmx}
\usepackage[T1]{fontenc}

\DeclareRobustCommand{\VAN}[3]{#2}
\let\VANthebibliography\thebibliography
\def\thebibliography{\DeclareRobustCommand{\VAN}[3]{##3}\VANthebibliography}

\usepackage{romannum}
\usepackage{graphicx}	% Including figure files
\usepackage{amsmath}	% Advanced maths commands
\usepackage{amssymb}	% Extra maths symbols

\usepackage{rotating}
\usepackage{longtable}
\usepackage[normalem]{ulem}
\useunder{\uline}{\ul}{}
\usepackage{csquotes}
\usepackage{multirow}
\usepackage{pdflscape}
\usepackage{supertabular}  
\usepackage{color, colortbl}
\newcommand{\code}{\texttt}
\definecolor{gray1}{gray}{0.8}
\definecolor{gray2}{gray}{0.8}
\definecolor{gray3}{gray}{0.8}
\cellcolor{gray3}
\cellcolor{white}

\title[Deep Learning in Searching the Spectroscopic Redshift of Quasars]{Deep Learning in Searching the Spectroscopic Redshift of Quasars}

\author[F.~Rastegar Nia]{
F.~Rastegar Nia,$^{1,2,3}$\thanks{f.rastegarnia@alzahra.ac.ir}
M.~T.~Mirtorabi,$^{1,2}$\thanks{torabi@alzahra.ac.ir}
R.~Moradi,$^{2,3,4}$\thanks{ rahim.moradi@inaf.it}
A.~Vafaei.~Sadr,$^{5}$\thanks{Alireza.VafaeiSadr@unige.ch}
Y.~Wang$^{2,3,4}$\thanks{ yu.wang@uniroma1.it}
\\
$^{1}$
 Department of Physics,Faculty of Physics and Chemistry, Alzahra University,Tehran, Iran\\
$^{2}$
ICRANet, Piazza della Repubblica 10, I-65122 Pescara, Italy\\
$^{3}$
ICRA, Dipartimento di Fisica, Universit\`a  di Roma ``La Sapienza'', Piazzale Aldo Moro 5, I-00185 Roma, Italy\\
$^{4}$
INAF -- Osservatorio Astronomico d'Abruzzo,Via M. Maggini snc, I-64100, Teramo, Italy\\
$^{5}$
Departement de Physique Theorique and Center for Astroparticle Physics, University of
Geneva\\
}

\date{Accepted XXX. Received YYY; in original form ZZZ}

\pubyear{2022}

\begin{document}
\label{firstpage}
\pagerange{\pageref{firstpage}--\pageref{lastpage}}
\maketitle

% Abstract of the paper
\begin{abstract}
Studying the cosmological sources at their cosmological rest-frames is crucial to track the cosmic history and properties of compact objects. In view of the increasing data volume of existing and upcoming telescopes/detectors, we here construct a 1--dimensional convolutional neural network (CNN) with a residual neural network (ResNet) structure to estimate the redshift of quasars in Sloan Digital Sky Survey IV (SDSS-IV) catalog from DR16 quasar-only (DR16Q) of eBOSS on a broad range of signal-to-noise ratios, named \code{FNet}.  Owing to its $24$ convolutional layers and the ResNet structure with different kernel sizes of $500$, $200$ and $15$, FNet is able to discover the ``\textit{local}'' and ``\textit{global}'' patterns in the whole sample of spectra by a self-learning procedure. It reaches the accuracy of 97.0$\%$ for the velocity difference for redshift, $|\Delta\nu|< 6000~ \rm km/s$ and 98.0$\%$ for $|\Delta\nu|< 12000~ \rm km/s$. While \code{QuasarNET}, which is a standard CNN adopted in the SDSS routine and is constructed by 4 convolutional layers (no ResNet structure), with kernel sizes of $10$, to measure the redshift via identifying seven emission lines (\textit{local} patterns), fails in estimating redshift of $\sim 1.3\%$ of visually inspected quasars in DR16Q catalog, and it gives 97.8$\%$ for $|\Delta\nu|< 6000~ \rm km/s$ and 97.9$\%$ for $|\Delta\nu|< 12000~ \rm km/s$. Hence, FNet provides similar accuracy to \code{QuasarNET}, but it is applicable for a wider range of SDSS spectra, especially for those missing the clear emission lines exploited by \code{QuasarNET}. These properties of \code{FNet}, together with the fast predictive power of machine learning, allow \code{FNet} to be a more accurate alternative for the pipeline redshift estimator and can make it practical in the upcoming catalogs to reduce the number of spectra to visually inspect.
\end{abstract}


%%%%%%%%%%%%%%%%%%%%%%%%%%%%%%%%%%%%%%%%%%%%%%%%%%

%%%%%%%%%%%%%%%%% BODY OF PAPER %%%%%%%%%%%%%%%%%%

\section{Introduction}

Quasi-stellar radio sources (Quasars) or quasi-stellar objects (QSO) are high-luminosity active galactic nuclei (AGN) which are well accepted to be powered by a gaseous accretion disk around a supermassive black hole (SMBH) with masses in the range of $\sim 10^6 M_{\odot}$ to $\sim 10^9 M_{\odot}$ \citep{2011Natur.474..616M,2019MNRAS.487.2030L}. Thanks to their high luminosity, quasars have been found to spread from redshift z$\sim$0 back to z$\sim$7 when the universe was forming its first structures, namely the epoch of reionization \citep{2020ApJ...897L..14Y,2020ARA&A..58...27I,2021ApJ...907L...1W}. Consequently, study the high-redshift quasars can be taken into account as a powerful tool to study the cosmic history and structure formation in the early universe; \citep[see e.g.][and references therein]{1993ARA&A..31..473A,2006AJ....131.1203F, 2018Natur.553..473B,2019MNRAS.488.4004L}. 

Recent studies show that quasars, owing to their existence in a wide range of redshift, provide a novel standard candle, like type Ia supernovae, by which new cosmological constraints to study the evolution of the universe can be inferred \citep{2020FrASS...7....8L}. For example, \citet{2019NatAs...3..272R} have recently shown that the deviation from the \textit{$\Lambda$}CDM inferred from type Ia supernovae emerges at z$>$1.4 with $\sim 4 \sigma$ statistical significance.

Moreover, quasars are perfect tools to study the physics governing the formation of SMBHs (the rapid growth of SMBHs occurs at the high redshift $z = 5\sim10$) and their surrounding accretion disk; see e.g., \citet{1993MNRAS.263..168H,2001ApJ...551L..27M,2010AJ....140..546W,2020ApJ...891...69C,2021A&A...649A..75M}.

On the one hand, the exploitation of spectroscopic data to utilize in physical studies needs an exact classification and redshift determination of astrophysical objects. On the other hand, the advancements of observational detectors have led to the immense growth of astronomical data. This massive volume of data produced in astrophysical surveys prevents the procedure for visual inspection of each spectrum. For example, the Sloan Digital Sky Survey IV (SDSS-IV) quasar catalog from Data Release 16 (DR16) of the extended Baryon Oscillation Spectroscopic Survey (eBOSS) \citep{2016AJ....152..205H,2020ApJS..250....8L}, includes a ``superset'' of objects targeted as quasars containing $1,440,615$ observations. Thus, implementing the \textit{automatic methods} which have the \textit{human--expert} precision are of great importance.

The richness of the data has brought huge opportunities for astronomers to develop intelligent tools and interfaces, utilizing the pipeline classifiers, the machine learning (ML), and the deep learning (DL) methods, to deal with data sets and extract novel information \citep{ball2010data,allen2019deep}.

In recent years several non-ML/DL redshift estimators, such as the principal component analysis (PCA) \citep{glazebrook1998automatic}, the automated redshift-finder by $\chi^2$ minimization in SDSS-III and SDSS-IV \citep{2012AJ....144..144B,2017A&A...597A..79P,2020ApJS..250....8L} as well as, pipeline classifier \code{redrock} in the upcoming Dark Energy Spectroscopic Instrument (DESI) quasar survey \citep{2020JCAP...11..015F}, have been introduced.

These standard automatic methods (template classifiers) work based on comparing each spectrum with a dataset of spectra. For example, in the $\chi^2$ minimization redshift-finder in SDSS-III \citep{2012AJ....144..144B}, each observed spectrum is linearly fitted and the output redshift is obtained for minimum $\chi^2$ when compared to specific templates. Also,  the PCA method which is included in DR16Q is a generalized cross-correlation method which instead of the individual templates uses a linear combination of orthogonal templates to summarize the information which exist in each spectrum \citep{glazebrook1998automatic}. These methods usually perform worse than human--expert methods in classification \citep{2013AJ....145...10D,paris2017sloan,2018arXiv180809955B}, while the DL/ML methods, thanks to their recognition patterns which identify spectral features, such as emission/absorption lines, spectral breaks, etc, perform the tasks as accurate as visual inspection level;  \citealp[see e.g.][]{2018arXiv180809955B}.

The template classifiers are typically used in conjunction with ML/DL methods to have an enhanced precision of classification and redshift estimation, for example, \citep{2020JCAP...11..015F} have shown that by combining the outputs of the \code{QuasarNET} \citep{2018arXiv180809955B} and the \code{redrock}, the classification results to identify the high redshift quasars from single exposures are improved and final quasar catalog contamination is reduced.

Over the last few years, ML and DL have become increasingly popular in astronomy and astrophysics to deal with big data surveys. DL/ML aims to seek and recognize, by the optimization procedure, all available common characteristics and patterns in data, which helps in turn to solve unseen problems \citep{RevModPhys.91.045002}. The redshift or any physical characteristics of the distant sources, consequently, can be estimated by utilizing DL networks recognizing deep patterns hidden in data. Some of these patterns probably are unknown or inapplicable by the traditional methods or even understanding of human beings. It is therefore expectable that DL methods can be extended to examine the spectra observed in various energy bands via various satellites like X-ray by Swift-BAT \citep{2017ApJS..233...17R} and gamma-ray by Fermi-LAT \citep{2020ApJ...892..105A}. 

These methods have been widely used for a variety of tasks, including morphological classification of galaxies \citep{dobrycheva2017machine,gauci2010machine,de2004machine}, evaluation of photometric redshift \citep{hoyle2016measuring,cavuoti2015machine, sadeh2016annz2,2018A&A...611A..97P}, star/galaxy classification \citep{kim2016star,odewahn1992automated,bai2018machine}, stellar spectra classification \citep{bailer1998automated,sharma2020application}, stellar atmospheric parameters estimation \citep{li2017parameterizing, fiorentin2007estimation}, analysis of stellar spectra \citep{fabbro2018application,bialek2019deep}, classification of quasars spectra and evaluating their redshift (SQUEzE) \citep{2020MNRAS.496.4931P,2020MNRAS.496.4941P}, photometric redshift estimation of quasars \citep{2018A&A...611A..97P}, and spectroscopic classification and redshift prediction of the quasars, as expertly as human visual inspection level, from the most confident emission lines, as the local properties of quasar spectra, e.g. \code{QuasarNET}  \citep{2018arXiv180809955B} which is adopted in constructing the DR16 catalog.

In this regard, in this paper we implement the \code{FNet}, a convolutional neural network (CNN) with a residual neural network (ResNet) structure to be trained by the observed optical flux of quasars, to estimate their redshift in Sloan SDSS-IV quasar catalog from DR16 of the eBOSS, which is the most comprehensive catalog of spectroscopically collected quasars to date \citep{2016AJ....152..205H,2020ApJS..250....8L}. Over 700,000 quasars present in the DR16Q catalog, with $326,535$ visually inspected (VI) quasars, makes it one of the best samples to invest the artificial intelligence methods. 

Both \code{FNet} and \code{QuasarNET} are CNN based networks, but with completely different designs. \code{QuasarNET} consists of 4 convolutional layers and kernel sizes of 10 to follow the traditional procedure to identify emission lines, as the local patterns hidden in spectra,  by its ``line finder'' units.  Instead \code{FNet}, consists of $24$ convolutional layers with a ResNet structure and kernel sizes of $500$, $200$ and $15$, finds the ``\textit{local}'' and ``\textit{global}'' patterns in spectra by a self-learning procedure. This makes the \code{FNet} applicable in the ambiguous spectra as well as a larger range of redshift; see section~ \ref{sec:results} for more details.

In section~\ref{sec:Data}, we briefly review the basics of quasar spectroscopy, the structure of their optical spectra as well as the description of SDSS and eBOSS surveys.

In section~\ref{sec:CNN} we provide the description of CNN structure developed for this work. 

In section~\ref{sec:results} present the results of training. We also present the comparison of our results with similar works in the literature. 

Finally, in section \ref{sec:conclusions} we represent the conclusion of the paper.

\begin{figure*}
\centering
(a)\includegraphics[width=0.48\hsize,clip]{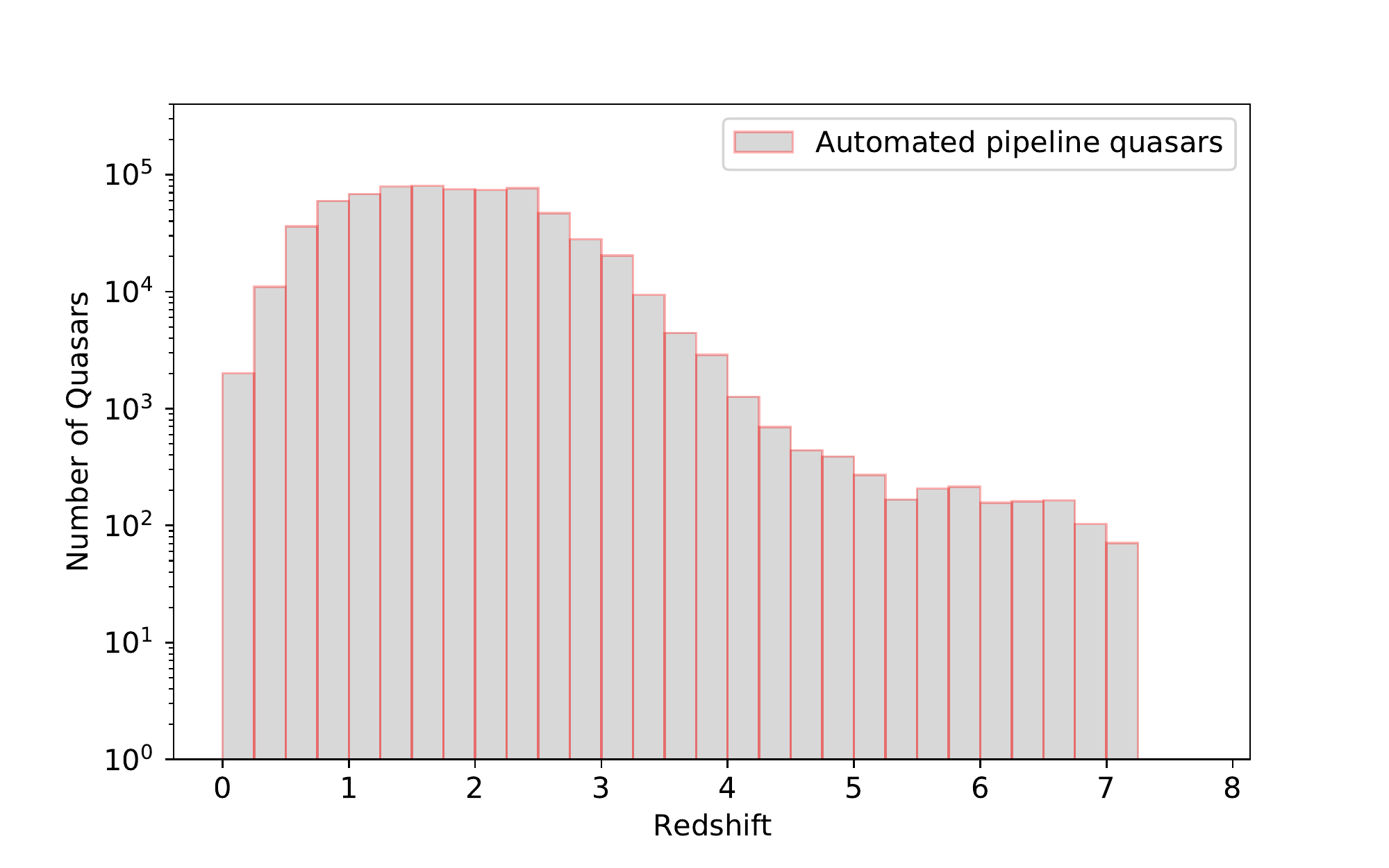}
(b)\includegraphics[width=0.48\hsize,clip]{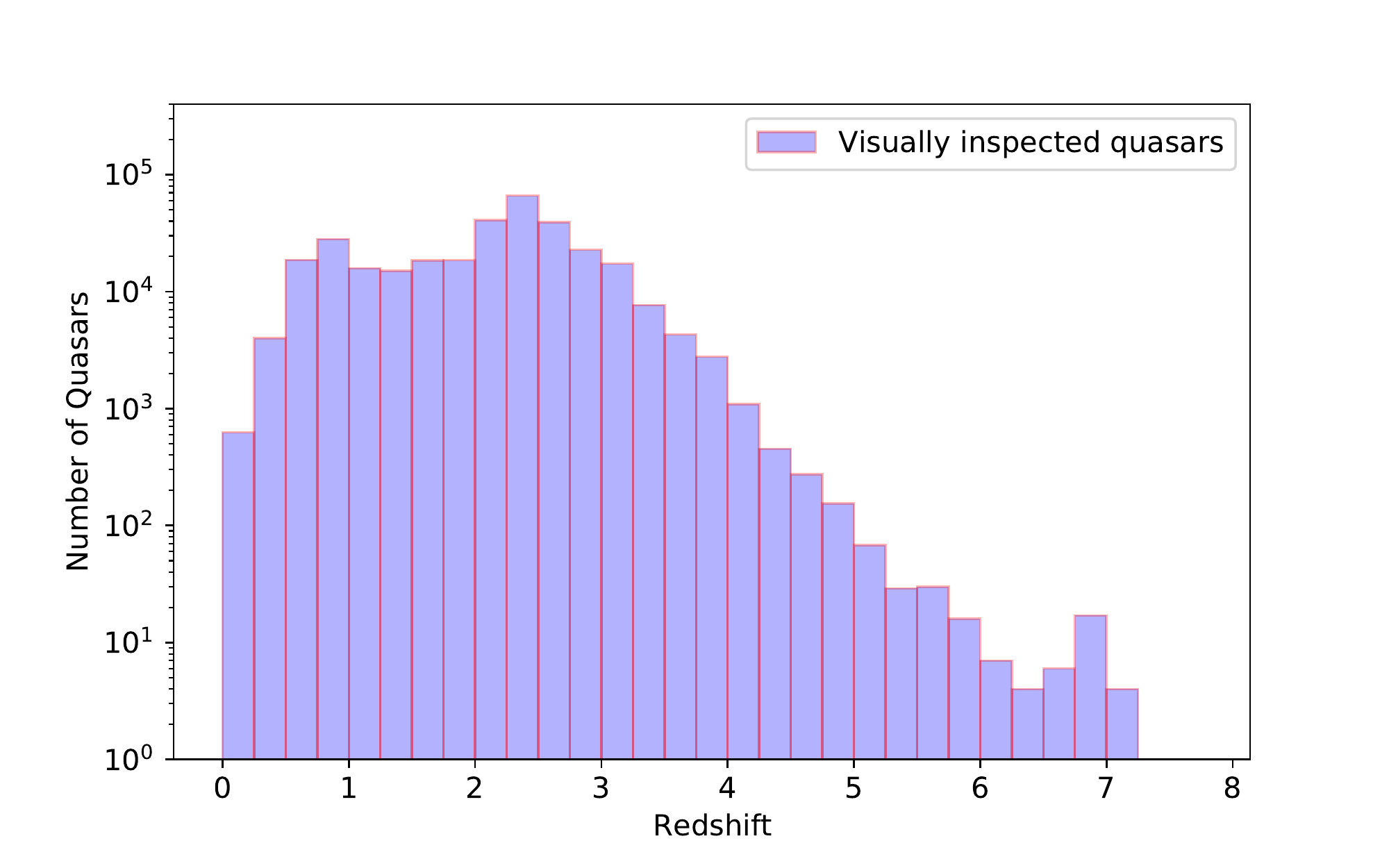}

\caption{The distributions of detected quasars redshift in the DR16 catalog for (a) pipeline quasars (750,414) as well as (b) the visually inspected quasars (326,535).  To have a sample with reliable quasars with human-expert precision, which is essential to train any neural network, only $326,535$ spectra of quasars with human-expert visually inspected (VI) classifications and redshift determinations are used in this work. }
\label{fig:histogram}
\end{figure*}
%

%%%%%%%%%%%%%%%%%%%%%%%%%%%%%%%%%%%%%%%%%%%%%%%%%%%%%%%%%
\section{Data}\label{sec:Data}
%%%%%%%%%%%%%%%%%%%%%%%%%%%%%%%%%%%%%%%%%%%%%%%%%%%%%%%%%

It is very well known that due to the expansion of the Universe and the subsequent cosmological redshift, $z \equiv \frac{{\lambda_{\rm obs}}-{\lambda_{\rm em}}}{\lambda_{\rm em}}$, the observed wavelength, $\lambda_{\rm obs}$, of the photons emitted from the quasar become longer than the emitted rest-frame wavelength ($\lambda_{\rm em}$). The spectroscopic analysis indicates that emission/absorption lines from quasars are significantly different from the spectra of stars. The spectra of quasars contain bright UV flux, broad emission lines, and often time-variable flux both in the continuum and in the emission lines;  \citep[see e.g.~][]{1996ima..book.....C}. This richness of spectra of quasars makes them proper tools to implement the DL method and extract the common features hidden in their spectrum (flux). 

The most comprehensive observed quasar (QSO) spectra to date are cataloged in the Sloan Digital Sky Survey-\Romannum{4}(SDSS-\Romannum{4}). SDSS has been operative from 2000 and created and released catalogs of quasars from 2002 \citep{2002AJ....123..567S}. The latest release of SDSS is provided by the Sixteenth Data Release Quasar-only (DR16Q) of SDSS extended Baryon Oscillation Spectroscopic Survey (eBOSS) \cite{dawson2016sdss,2020ApJS..250....8L}. In eBOSS, data are recorded from 500 fibers on a 2k CCD related to each spectrograph and, the wavelength ($\lambda$) coverage is in the range of $\sim 361$--$1014$~nm \citep{2020ApJS..250....8L}.

DR16Q comprises automated classifications and redshifts determined by version \code{v5\_13\_0} of the SDSS spectroscopic pipeline \citep{2020ApJS..250....8L}.  This catalog contains 750,414 quasars, with the automated redshift range $0 \leq z \leq 7.1$.  The distributions of quasars in terms of their pipeline redshift in the DR16Q catalog is shown in Fig.~\ref{fig:histogram} (a). As it can be seen, the number of sources reaches its maximum around $z\approx2.5$; at earlier epochs i.e., higher redshifts, they are comparatively rare.

The problem with SDSS-DR16Q catalog is that due to wrong pipeline classification and redshift estimation, it contains sources that are shown to be non-quasars. This catalog is contaminated with 0.3$\%$–1.3$\%$ of non-quasar objects \citep{2020ApJS..250....8L}. For example, in a search for undeclared quasars,  \citet{2021MNRAS.tmp..804F} has shown that 81 entries in the SDSS-DR16Q main quasar are not quasars. Therefore, the pipeline catalog is not an adequate training sample for quasars, especially for the quasars with z$>$4, as many of the listed objects with z$>$6, and potentially even substantial fractions of objects at z$>$4, are not quasars or not quasars at the given redshifts due to incorrect pipeline classifications and redshifts from SDSS. 

Therefore, to obtain a sample with reliable quasars and human-expert precision, which is essential to train any neural network, we here constrain data on $326,535$ spectra of quasars with human-expert visually inspected (VI) classifications and redshift determinations. The quasars are annotated by QSO; \code{CLASS\_PERSON}= 3 in DR16Q catalog \citep{2020ApJS..250....8L}. The distributions of VI quasars in the DR16Q catalog is shown in Fig.~\ref{fig:histogram} (b) which are approximately half of the total numbers of quasars in this catalog.

This sample, in addition to the visually inspected QSOs from DR7Q \citep{2010AJ....139.2360S} and DR12Q \citep{2017A&A...597A..79P}, includes the ill-identified QSOs from DR14Q \citep{2018A&A...613A..51P} and DR16Q catalogs, which were flagged for visual inspection when the automated methods fail to identify them as QSO. Most of these QSOs correspond with a low S/N ratio or show strong absorption lines which confuse the pipeline; see section.~3 of \cite{2018A&A...613A..51P} and section.~3 of \citet{2020ApJS..250....8L}. This is strong evidence to demonstrate why testing any net only on DR12Q provides more accurate results than the one of visually inspected DR16Q.

Therefore, for the sake of (1) maximizing the reliability and uniformity of the training sample and, (2) comparison with \code{QuasarNet}, we use the training sample used by \code{QuasarNet}, which includes 249,762 unique sources flagged as QSO in DR12Q catalog, as our training set. The spectra in the test set are chosen among the VI quasars included in the DR16Q catalog included to the catalog after the DR12Q catalog release date. In the whole paper, the test set has not been used to train the CNN.

\section{Convolutional neural network (CNN)}\label{sec:CNN}

\begin{figure}
\centering
\includegraphics[width=1.0\hsize,clip]{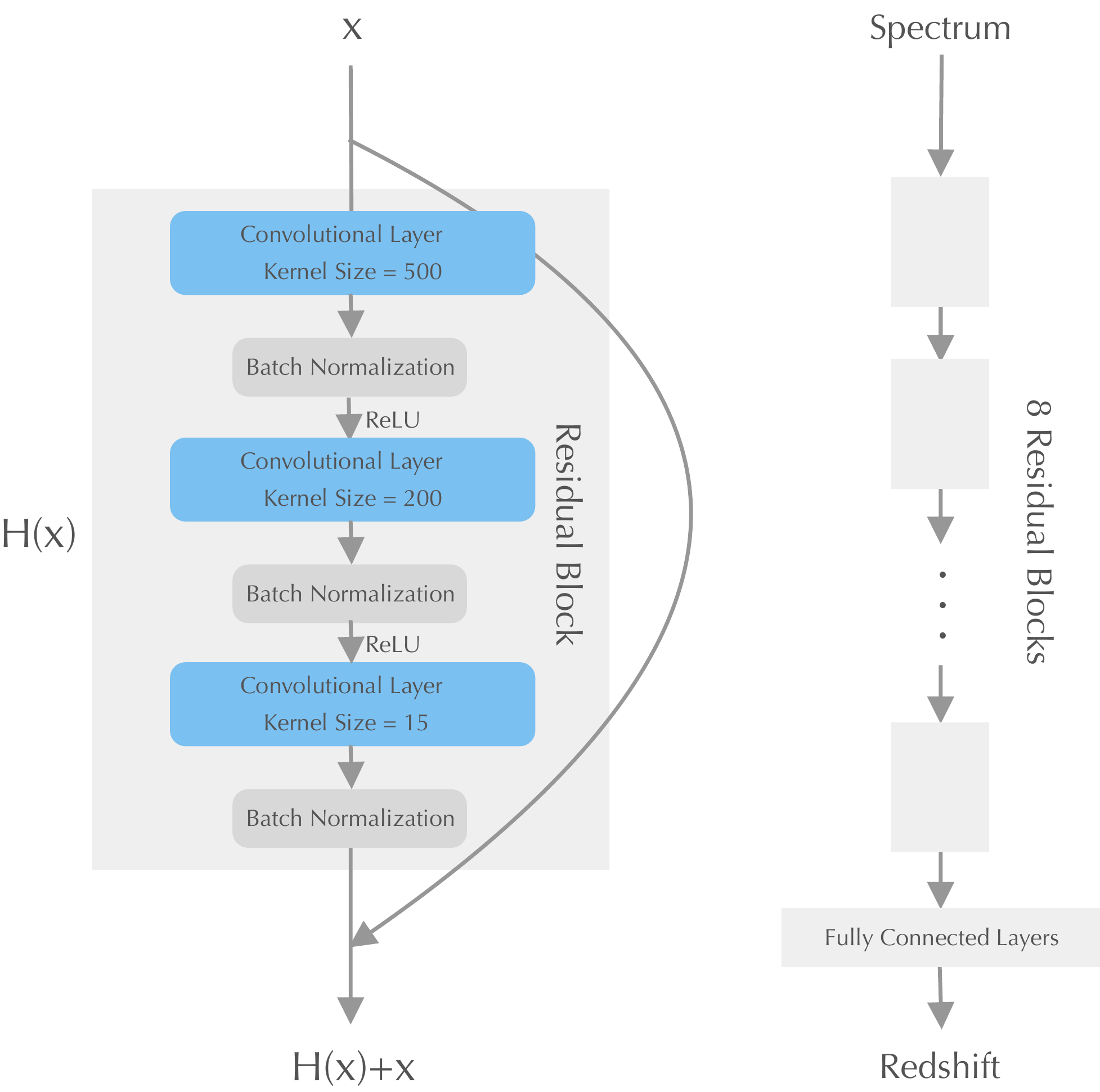}
\caption{The chosen architecture of 1-dimensional CNN developed in this work (\code{FNet}) to  learn higher-order features hidden in the input flux. The CNN slides the flux via convolutional layers of kernel size = 500, 200 and 15, respectively to search for the ``global'' and ``local'' patterns in the flux of quasars. The fully connected layers output the redshift. \textbf{Left}: The structure of a residual block, the input $x$ goes through two convolutional layers as $H(x)$ then add itself as $H(x)+x$, batch normalization is applied after each convolutional layer, and the activation function ReLU acts on the first batch normalization layer. \textbf{Right}: The entire structure: The flux goes through $24$ residual blocks, the first $21$ blocks have channel size $32$, followed by three blocks of channel size $64$, $32$ and $16$ respectively. The output of blocks then is flattened and passes three fully convolutional layers and eventually outputs the redshift; ReLU is applied after each fully connected layer.} 
\label{fig:CNN-structure}
\end{figure}
%
%%%%%%%%%%%%%%%%%%%%%%%%%%%%%%%%%%%%%%%%%%%%%%%%%%%%%%%%%
\subsection{preprocessing the data}
%%%%%%%%%%%%%%%%%%%%%%%%%%%%%%%%%%%%%%%%%%%%%%%%%%%%%%%%%

Data preprocessing plays an important role to provide understandable inputs for DL networks. An astronomical object's spectrum is similar to a time series since it is 1+1 dimensional where each data point presents wavelength versus the flux. Each raw spectrum in DR16Q contains nearly 4500 data points of flux situated in the logarithmic space of wavelength. We standardize each spectrum, by fitting and extrapolating, to a 1-dimensional vector consisting of 4618 data points (pixels) uniformly situated in  $\log \lambda$, being $\lambda$  the wavelength in the range of $360~\rm nm-1032.5~\rm nm$. We then normalize the fluxes of each spectrum via the Zero-mean normalization and the Unit-norm normalization \citep{jayalakshmi2011statistical}. Due to the existence of large variations in the distribution of quasar flux which makes the training process slow and unstable, this step is necessary for the sake of accelerating the network and increasing the accuracy of processing. These reduced, normalized fluxes that are used to feed the neural network are stored in $N\times M$ matrix, in which $N$ is the number of quasars to be studied and $M$ is the number of pixels. On the other hand, normalized redshift is applied as labels which are similarly stored in a $N\times 1$ matrix, in which $N$ is the number of quasars. 

%%%%%%%%%%%%%%%%%%%%%%%%%%%%%%%%%%%%%%%%%%%%%%%%%%%%%%%%% 
 \subsection{CNN architecture}
%%%%%%%%%%%%%%%%%%%%%%%%%%%%%%%%%%%%%%%%%%%%%%%%%%%%%%%%%

The Convolutional Neural Networks (CNNs) have been used widely for machine learning tasks \citep{Goodfellow-et-al-2016}.
They have shown considerable improvements in both computer vision \citep{xu2014deep, koziarski2017image, yamashita2018convolutional} and time-series problems \citep{yang2015deep,liu2018time}. In this work, 1-dimensional CNN of ResNet structures \citep{he2016deep} is designed to find the higher-order features in the input observed flux. The optimized convolutional layers transform the features to the visually inspected redshift in the output layer. In this view, the neural network model is only a mathematical function that maps an input (observed flux) to the desired output (redshift). 

As such, our \code{FNet} is similar to \code{QuasarNET} \citep{2018arXiv180809955B} which is also a CNN based network, however, the design thinking is completely different. \code{QuasarNET} imitates the traditional procedure of redshift estimation, namely first to identify the prominent lines, then to infer the redshift from the detected lines. In our CNN, instead, the net is trained to discover the hidden patterns by itself without given any external information including lines. These hidden patterns, in addition to prominent emission lines targeted by \code{QuasarNET}, can be other emission/absorption lines present in spectra,  global shifted patterns, specific patterns at different redshifts, or patterns related to the correlation of fluxes at different wavelengths. These two approaches of thought lead to different results and a different range of applications, which we address in the article.

In addition, to find more abstract patterns and to have more accurate predictions, \code{FNet} is deepened by incorporating $24$ convolutional layers in its architecture. From our current knowledge of neural networks  \citep{albawi2017understanding, aloysius2017review, sengupta2020review}, each layer of the neural network extracts different levels of feature information. The deeper the network, the more information will be extracted. This strategy strengthens capabilities of \code{FNet} to generalize its prediction for spectra with a wide range of signal-to-noise ratios, as well as spectra without prominent emission lines.

However, the deep networks always encounter the problem of gradient vanishing or exploding, we therefore adopt the techniques of batch normalization \citep{ioffe2015batch} and gradient clipping \citep{pascanu2013difficulty} to maintain the gradients of all layers in a appropriate range. Deep networks also face the problem of gradient degradation, that the accuracy drops when the number of layers are increased. To tackle this problem,  the residual structure is incorporated \citep{he2016deep}. We stack three convolutional layers as a residual block, then connect 8 blocks in series. In parallel, we implement shortcuts between blocks that sum up the input from the previous block and the output of the current block. This net structure optimizes the residuals $\rm H(x) - x$, where x is the input and $\rm H(x)$ is the effective map of a block. If the residual is zero, $\rm H(x)$ does an identity mapping. If the residual is not zero, $\rm H(x)$ learns new features. Hence, this residual structure prevents the degradation problem which worsens the net performance.

In \code{FNet}, each convolutional layer has $32$ or $64$ filters (channels) sliding along each row (flux) to extract the prominent features which will be passed to a Rectified Linear Unit (ReLU) activation function that allows the models to access non-linear modes if they are needed \citep{xu2015empirical}. The convolutional layers of a small filter (kernel size = 15), medium filter (kernel size = 200) and a large filter (kernel size = 500) are stacked as a residual block and duplicated in series forming a deep network of in total $24$ convolutional layers, shortcuts are applied between each two residual blocks; see Fig.~\ref{fig:CNN-structure}. It should be pointed out that the convolutional layers are initialized with He Normal initializer \citep{he2015delving}. Finally, the fully-connected feed-forward layers connect and assign the extracted features to the output (redshift). All free parameters in the model change dynamically as the algorithm finds the best solution, achieved by the back-propagation learning algorithm.

The model architecture, as the layer specifications and their arrangement, is selected through the procedure of hyper-parameter optimization \cite{feurer2019hyperparameter}. We have tested the efficiency of training and the accuracy of prediction using different filters sizes (from 10 pixels to 500 pixels) and model depths (from 8 layers to 72 layers). The chosen architecture in figure \ref{fig:CNN-structure} reaches an optimized balance of efficiency and accuracy. 

\begin{figure}
\centering
\includegraphics[width=1.0\hsize,clip]{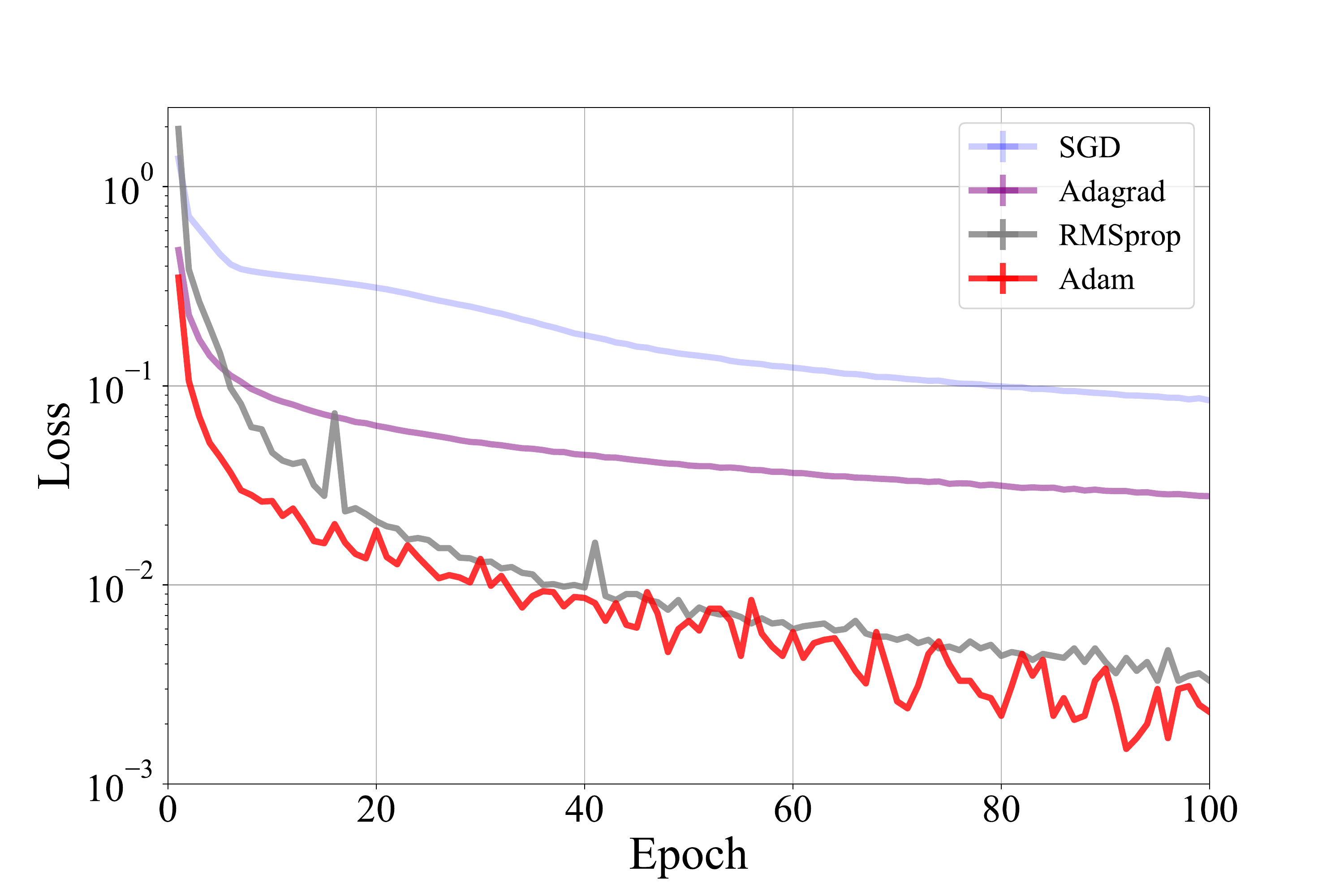}
\caption{The loss decay (in log scale) for \code{Adam}, \code{SGD},\code{AdaGrad} and \code{RMSProp} optimizers obtained for 100 epochs. Clearly, using \code{Adam} optimizer, results in a faster decay of loss function and consequently more accurate prediction. The train is performed over 63000 spectra; 90\% train and 10\% test. The accuracy for $|\Delta \nu|<$6000 km/s ( Eq.~\ref{eq:deltav}) is \code{Adam}: 97\%, \code{SGD}: 30\%, \code{AdaGrad}: 60\%  and \code{RMSProp}: 95.8\% . The learning rate for this comparison is 0.00024 for all optimizers. No step learning is implemented in this test.}  
\label{fig:comarisions}
\end{figure}

One also needs to choose the optimization method. In this work, the ``Mean Squared Error'' (MSE) is used as the loss function in this regression problem as well as  \code{Adam} optimizer to optimize the loss function. The \code{Adam} is an extension of stochastic gradient descent (\code{SGD}) which utilizes the first-order gradient-based optimization of stochastic objective functions to adjust the learning rate. This optimizer takes advantage  of two \code{SGD}--base optimizers: 1. Adaptive Gradient Algorithm; \code{AdaGrad} \citep{JMLR:v12:duchi11a}, mostly for sparse gradients, and 2. Root Mean Square Propagation; \code{RMSProp} \citep{Tieleman}, for on--line and non--stationary settings. \code{Adam} is well known in deep learning community due to its fast convergence to the optimized results comparing to \code{SGD}, \code{AdaGrad} and \code{RMSProp} \citep[see][for more details]{kingma2014adam}. We have compared the accuracy and the loss decay of \code{Adam} with \code{SGD},\code{AdaGrad} and \code{RMSProp} optimizers. For performing this test we trained only 63000 spectra of DR12Q catalog (90\% train and 10\% test) over 100 epochs. Results, clearly show the privilege of using \code{Adam} in the redshift prediction of quasars. The accuracy calculated for $|\Delta \nu|<$6000 km/s (defined by Eq.~\ref{eq:deltav} in Sec.~\ref{sec:results}) is \code{Adam}: 97\%, \code{SGD}: 30\%, \code{AdaGrad}:60\%  and \code{RMSProp}: 95.8\%; see Fig.~\ref{fig:comarisions}. 

 \begin{figure}
\centering
[a]\includegraphics[width=0.95\hsize,clip]{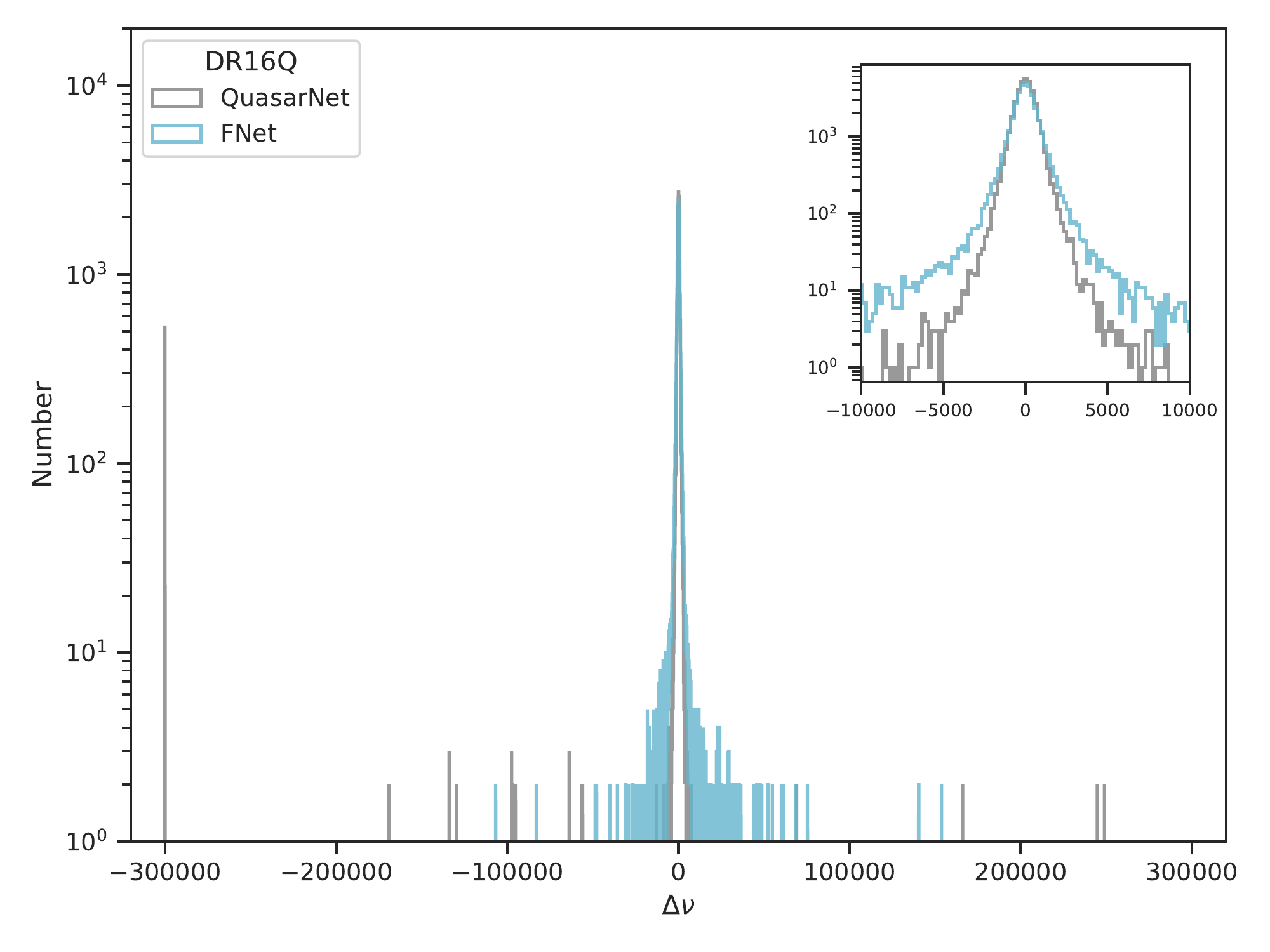}
[b]\includegraphics[width=0.95\hsize,clip]{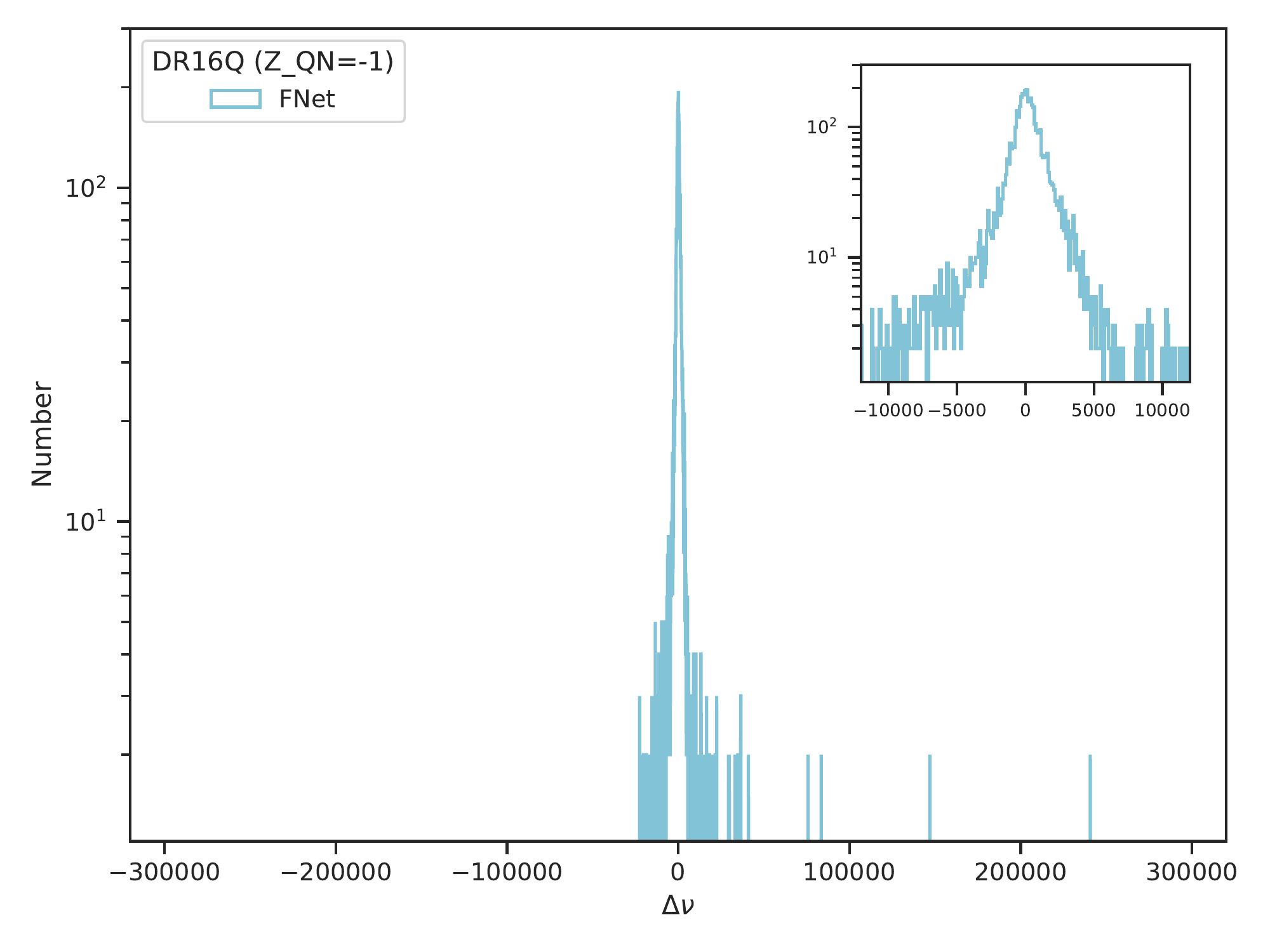}
[c]\includegraphics[width=0.95\hsize,clip]{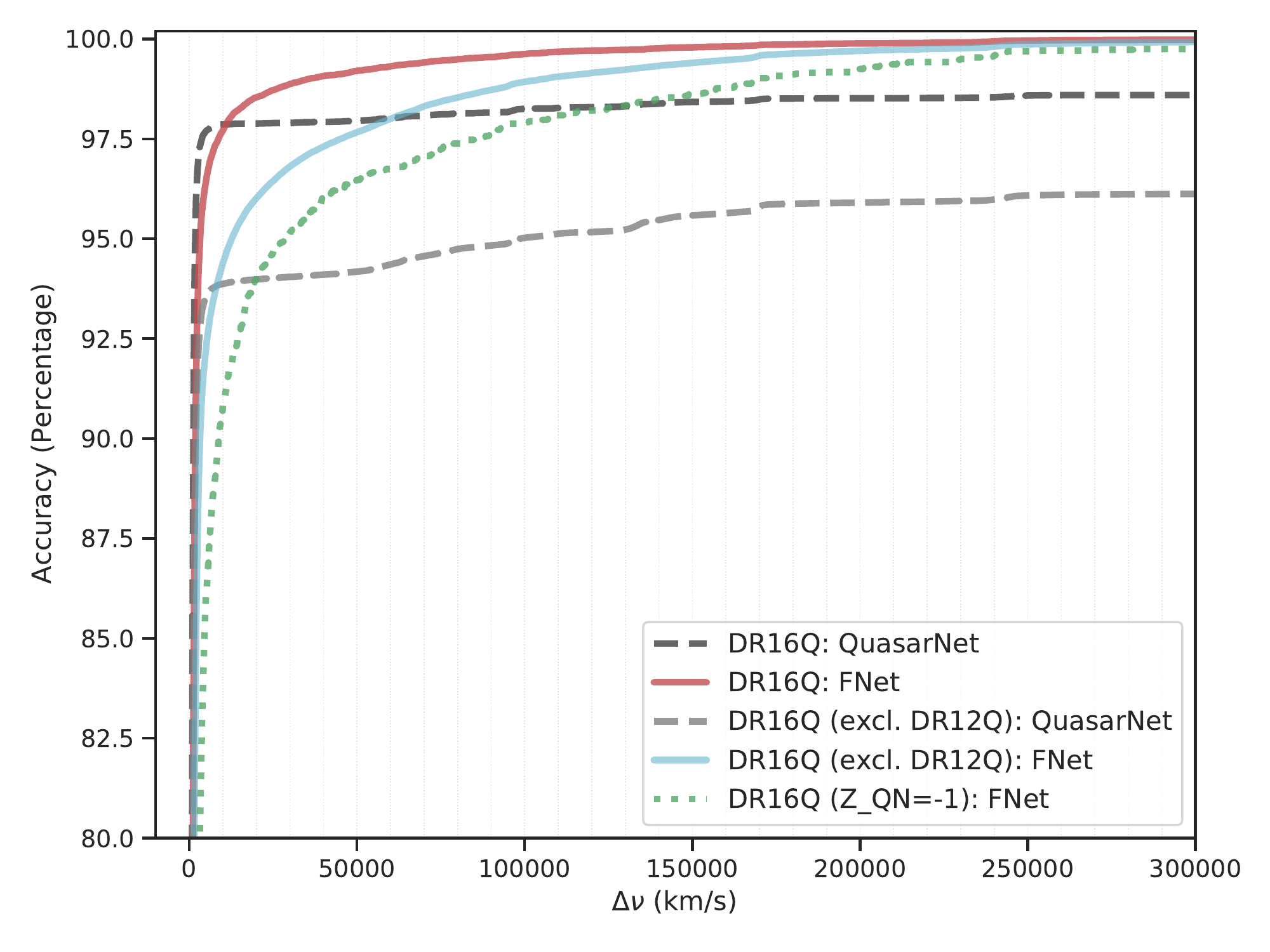}

\caption{\textbf{[a]}: The predicted redshift vs. the  visually inspected (VI) redshift, selected from the DR16Q catalog as explained in section~\ref{sec:Data} for \code{FNet} and \code{QuasarNET}. For $|\Delta \nu| < 6000 \rm~km/s$, the accuracy of \code{FNet} is 97.0$\%$ and for \code{QuasarNET}, 97.8$\%$. The accuracy for $|\Delta \nu| < 12000 \rm~km/s$ is 98.0$\%$ for \code{FNet} and 97.9$\%$ for \code{QuasarNET}.  \textbf{[b]:} The \code{FNet} redshift estimation for 5,190 visually inspected sources, flagged and reported as quasars in DR16Q, with 91.6\% accuracy for $|\Delta \nu| < 12000 \rm~km/s$. \code{QuasarNET} failed to predict this sample ($Z_{\code{QN}}=-1$ in DR16Q catalog).  \textbf{[c]}: The accuracy of prediction vs. $\Delta \nu$ for both \code{FNet} and \code{QuasarNET}. The red solid line and black dashed line represent the accuracy of the \code{FNet} and \code{QuasarNET}, respectively, for DR16Q sample. The blue solid line and grey dashed line represent the accuracy of the \code{FNet} and \code{QuasarNET}, respectively, for DR16Q sample when DR12Q is excluded. The green dashed lines shows the accuracy of \code{FNet} for  5,190 visually inspected sources in DR16Q when \code{QuasarNET} fails to estimate ($Z_{\code{QN}}=-1$.)   }  
\label{fig:Results}
\end{figure}

\begin{figure*}
\centering
\includegraphics[width=0.80\hsize,clip]{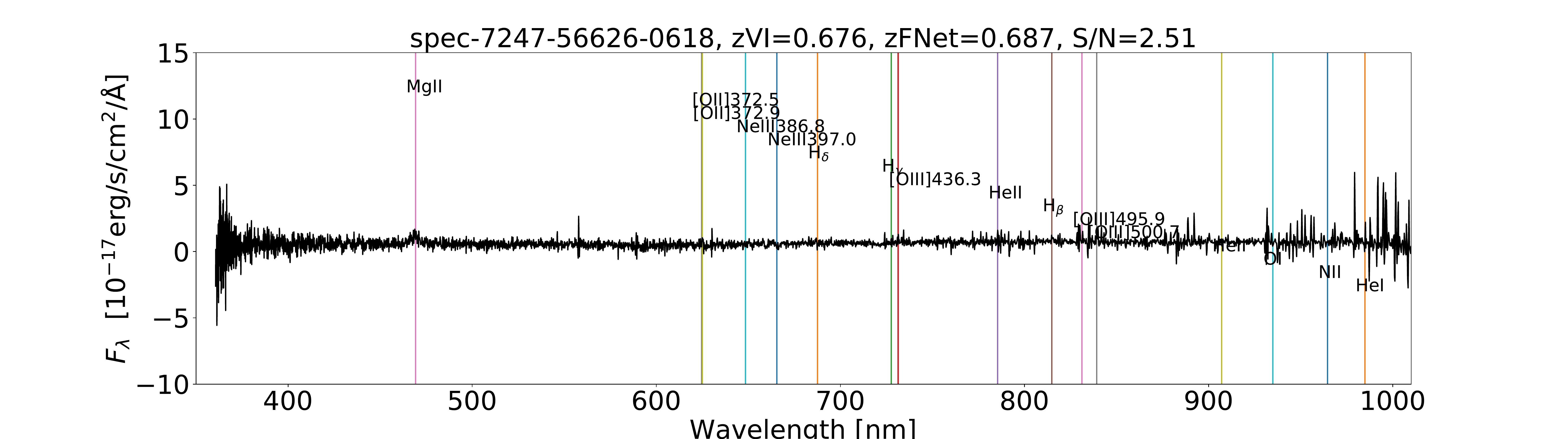}
\includegraphics[width=0.80\hsize,clip]{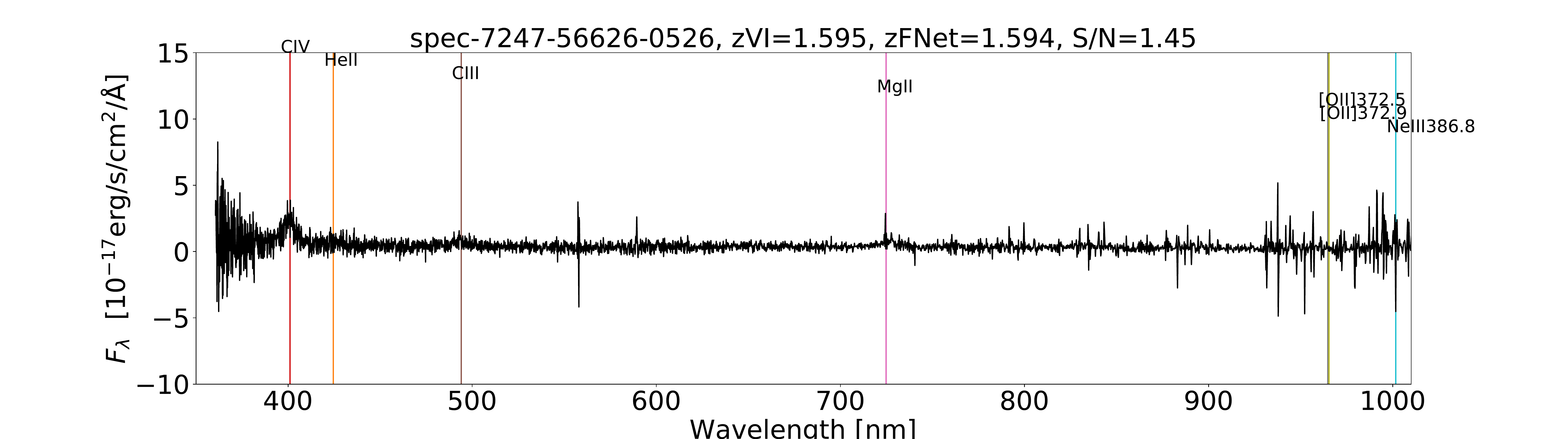}
\includegraphics[width=0.80\hsize,clip]{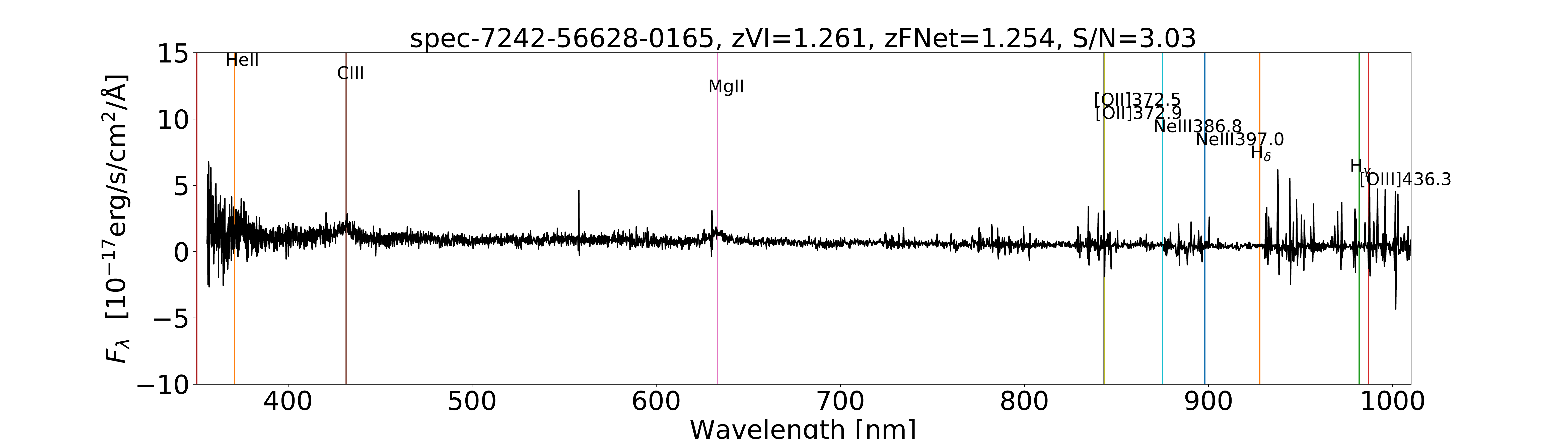}
\includegraphics[width=0.80\hsize,clip]{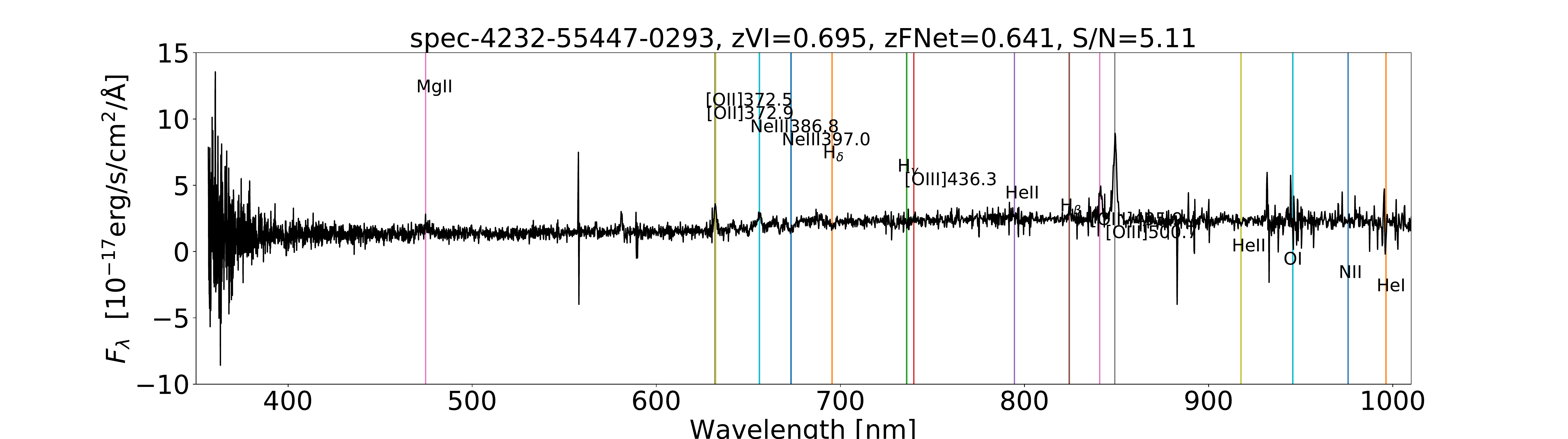}
\caption{The spectra of a quasars with failed \code{QuasarNET}  redshift ( $Z_{\code{QN}}=-1$) in VI catalog of DR16Q. Although there are several emisson/absorption lines present in the spectra, \code{QuasarNET} fails to find at least two emission lines among  Ly$\alpha$ (121.6 nm), $\rm CIV$ (154.9 nm), $\rm CIII$ (190.9 nm), $\rm MgII$ (279.6 nm), $\rm H\beta$ (486.2 nm) and $\rm H\alpha$ (656.3 nm) as well as a $\rm CIV$ line with a broad absorption feature, to be fed to its line finder units. In most of such quasars, due to low signal to noise no prominent emission lines  or at most one significant emission line among 7 emission lines to be detected by \code{QuasarNET} are present. \code{FNet} predicts the redshift of 5,190 such VI quasars with 87.4\% accuracy for $|\Delta\nu|< 6000~ \rm km/s$, 91.6\% accuracy for $|\Delta\nu|< 12000~ \rm km/s$ and 95.1\% accuracy for $|\Delta\nu|< 30000~ \rm km/s$.}  
\label{fig:failedQN}
\end{figure*}

Moreover, during the training progresses, the step learning rate is used, namely after some epochs the learning rate is reduced by an order of 0.1. We set the initial learning rate $\rm lr_1=0.00024$, after 75 epochs $\rm lr_2=0.1~lr_1$, after 85 epochs $\rm lr_3=0.01~lr_1$ and after 90 epochs $\rm lr_4=0.001~lr_1$.  The coefficients for computing the averages of gradient and its square was set as $\beta_1=0.9$ and $\beta_2=0.999$, the weight decay of L2 penalty as $1\rm e$--5 \citep{lecun1998gradient,kingma2014adam}. Traditionally methods like cross-validation are used to handle the overfitting problem with a small set of features present in the dataset. However, when a large number of features are present in the dataset, alternative methods like L2 penalty weight decay can be helpful.  L2 penalty, as it is obvious from its name, adds ``squared magnitude'' of weight as a penalty term to the loss function. In our case for training in the whole redshift range, we have checked the different weight decay values and weight--decay= $1\rm e$--5 leads to the most optimized result. In fact this small value of weight decay shows the almost equal importance of different features hidden in the dataset.

Figure~\ref{fig:CNN-structure} shows the chosen architecture of our CNN (\code{FNet}) developed in  this work which takes a quasar's fluxes as a 1-dimensional vector and predicts the redshift. First, in the training phase, the dataset with its labels is presented to the network. In this phase, the redshift is evaluated and network weights are adjusted to increase the accuracy and reduce the loss. Second, in the testing phase, the independent testing dataset is presented to the network. The percentage of data to allocate for training, validation and test sets to be fed to \code{FNet} is represented in the next section.

The training and testing of our network is performed with the \code{Torch} neural network library, which provides a high level application  program  interface  to  the \code{PyTorch} \citep{2019arXiv191201703P} machine intelligence software package. We take advantage of the \code{Skorch} \citep{skorch}, which is a library in \code{Pytorch} for machine learning models especially neural networks. It is a powerful tool that combines \code{Pytorch} and \code{Scikit-learn} \citep{2012arXiv1201.0490P}. The codes can be found in: \href{https://github.com/AGNNet/FNet.git}{https://github.com/AGNNet/FNet.git}, and the prepared data set can be found in: \href{https://www.kaggle.com/ywangscience/sdss-iii-iv}{https://www.kaggle.com/ywangscience/sdss-iii-iv}. We train \code{FNet} on a Nvidia Tesla V100 graphic card, training each epoch costs $90$~minutes. The time of inferring redshift of a SDSS spectrum costs $14$~ms.

\begin{figure*}
\centering
\includegraphics[width=1.0\hsize,clip]{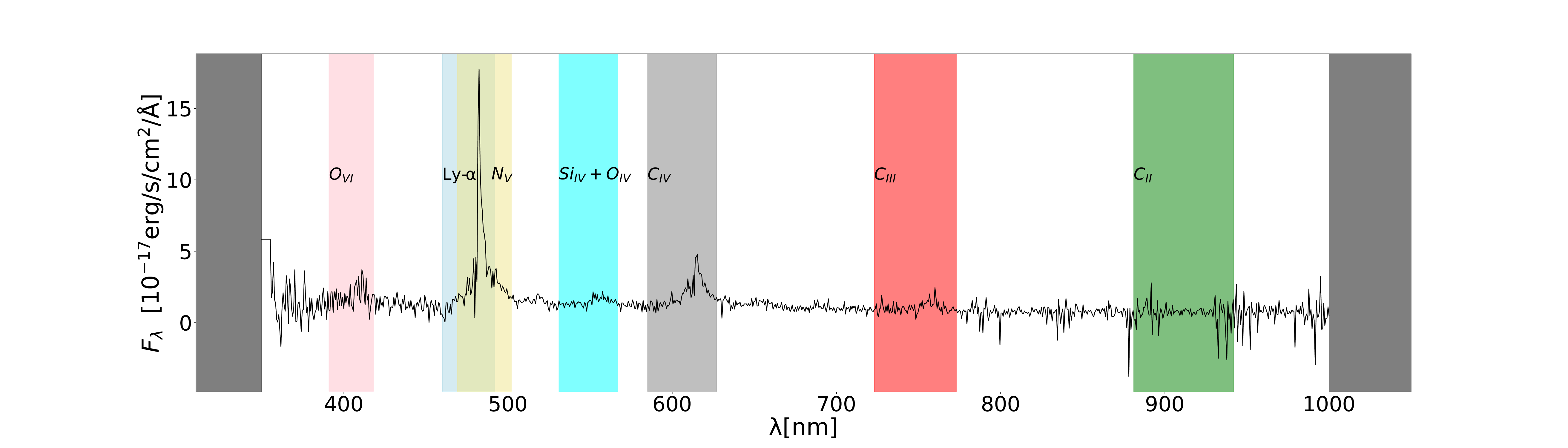}
\caption{The spectrum of a quasar with redshift of z=2.976. The color region shows the common wavelength interval removed from spectrum of all quasars in $2.8 \leq z \leq 3$ interval in order to remove the $\rm O{\rm VI}$ (103.3 nm, pink), Ly-$\alpha$ (121.6 nm, light blue), $\rm N{\rm V}$ (124.1 nm, khaki), $\rm Si{\rm IV}+OIV$ (139.8 nm, cyan),  $\rm C{\rm IV}$ (154.9 nm, grey) and $\rm C{\rm III}$ (190.9 nm, red), $\rm C{\rm II}$ (232.6 nm, green) lines from all quasars spectra in the aforementioned redshift interval.}  
\label{fig:2.8-3-mask}
\end{figure*}
\begin{figure}
\centering
\includegraphics[width=1.0\hsize,clip]{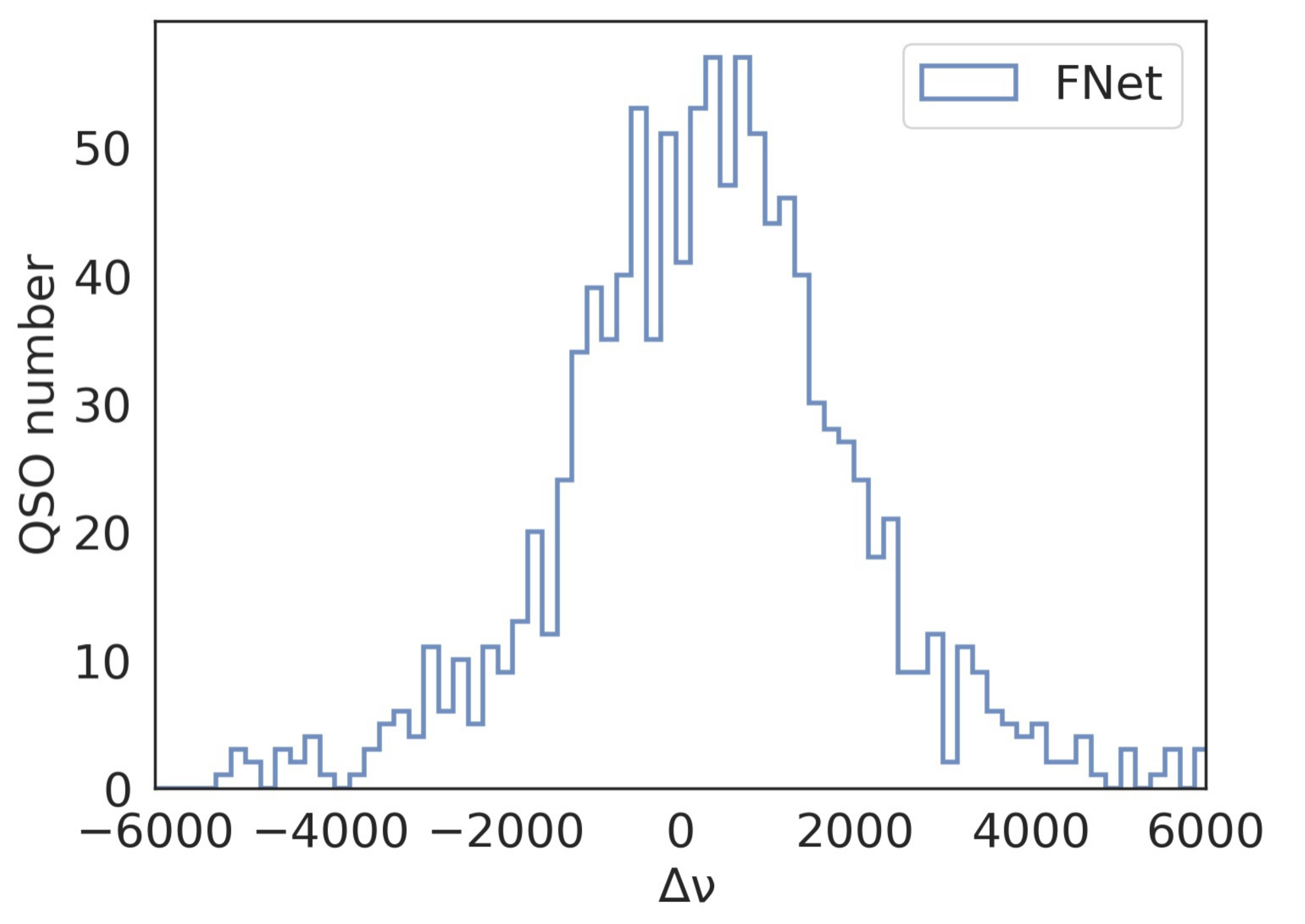}
\caption{The velocity difference for redshift predicted by \code{FNet} when compared to visually inspected (VI) redshift for $ 2.8 \leq z \leq 3$ interval. The $\rm O{\rm VI}$ (103.3 nm), Ly-$\alpha$ (121.6 nm,), $\rm N{\rm V}$ (124.1 nm), $\rm Si{\rm IV}+OIV$ (139.8 nm),  $\rm C{\rm IV}$ (154.9 nm) and $\rm C{\rm III}$ (190.9 nm), $\rm C{\rm II}$ (232.6 nm) lines are removed from all quasars spectra in the aforementioned redshift interval. The accuracy for \code{FNet} for $|\Delta \nu| < 6000 \rm~km/s$ is 98.5$\%$. The \code{QuasarNET} requires emission lines to estimate the redshift, therefore, applying this test on it would result in null prediction.}  
\label{fig:Results1}
\end{figure}
%

%%%%%%%%%%%%%%%%%%%%%%%%%%%%%%%%%%%%%%%%%%%%%
\section{results}\label{sec:results}
%%%%%%%%%%%%%%%%%%%%%%%%%%%%%%%%%%%%%%%%%%%%%

As discussed in section \ref{sec:Data} we use the entire $326,535$ visually inspected quasars of DR16Q catalog for training and testing \code{FNet}. 90\% of DR12Q (224786 spectra) are used as training set, 10\% of  DR12Q (24976 spectra) for validation by the procedure of k-fold cross-validation \citep{refaeilzadeh2009cross} and remaining visually inspected quasars of DR16Q (76773 spectra) are used as test set. The test set has not been used to train the CNN. Moreover, no limit on the S/N ratio, and no limit on the emission line availability are imposed. 
Following \citet{2020ApJS..250....8L} we define the velocity difference from redshift: 
\begin{equation}\label{eq:deltav}
    \Delta\nu = \rm c \times  \frac{Z_{\rm C}-Z_{VI}}{1+Z_{VI}}, 
\end{equation}
in which $c=2.998 \times 10^8 \rm~km/s$ is the speed of light, $\rm Z_{VI}$ is the redshift from the visually inspected quasar sample and $Z_{\rm C}$ is the redshift predicted by any model. The accuracy is defined as the percentage of samples contained within a given $|\Delta\nu$|.

The velocity difference for redshift predicated by \code{FNet}, when compared to visually inspected redshift is 97.0$\%$ accurate for $|\Delta \nu| < 6000 \rm~km/s$; see Fig.~\ref{fig:Results}. The accuracy is 98.0$\%$ for $|\Delta \nu| < 12000 \rm~km/s$ and 98.9$\%$ for $|\Delta \nu| < 30000 \rm~km/s$.
%

%%%%%%%%%%%%%%%%%%%%%%%%%%%%%%%%%%%%%%%%%%%%%%%%%%%%%%%%%
%\subsection{\code{QuasarNET} vs FNet}\label{subsec:QS-FN}
%%%%%%%%%%%%%%%%%%%%%%%%%%%%%%%%%%%%%%%%%%%%%%%%%%%%%%%%%
In order to compare our results and contrast the configuration of our CNN with other networks, in addition to our results, we here represent also the results presented by \citep{2018arXiv180809955B}.

Their network, known as \code{QuasarNET} is promising and provides high accuracy prediction in both quasar classification and redshift prediction, as competently as the human--expert confidence, which has already been exploited in DR16Q \citep{2020ApJS..250....8L} to significantly lower the number of quasars which need the human-expert visually inspection. 

The \code{QuasarNET} exploits a  convolutional neural network (CNN), consisting of 4 convolutional layers and one fully-connected layer as the final layer to feed into 7 number of ``line finder'' units. The filter size in each convolutional layer is 10, this network aims to find the local characteristics of spectra; namely emission lines. It trains a network by DR12Q data set which contains confirmed quasars via visual inspection of the spectra, to detect seven emission lines in the quasar's spectra: Ly$\alpha$ (121.6 nm), $\rm CIV$ (154.9 nm), $\rm CIII$ (190.9 nm), $\rm MgII$ (279.6 nm), $\rm H\beta$ (486.2 nm) and $\rm H\alpha$ (656.3 nm) as well as a $\rm CIV$ line with a broad absorption feature. \citet{2018arXiv180809955B} have selected the above lines such that at least two are present and detectable in the ``\textit{optical spectrograph of BOSS}''.

The accuracy of this network in predicting redshift in the range of 0$<$~z~$<$5.45 is 99.8$\%$ for $|\Delta\nu|< 6000~ \rm km/s$ with  $\rm Z_{VI}$ being the redshift obtained via visual inspection of the spectra in DR12Q catalog \cite{2017A&A...597A..79P}, as well as 97.8 ($|\Delta\nu|< 6000~ \rm km/s$),  97.9 \% ($|\Delta\nu|< 12000~ \rm km/s$) and 98.0 \% ($|\Delta\nu|< 30000~ \rm km/s$) for DR16Q catalog. As mentioned in section~\ref{sec:Data}, the presence of the ill-identified quasars, namely (1) the low S/N ones, or (2) the ones which show the presence of strong absorption lines in catalogs after DR12Q leads to the lower accuracy of the estimation of \code{QuasarNET} when applying on DR16Q.

Moreover, in DR16Q, as indicated by \citet{2020ApJS..250....8L}  0.6\% (8,581 spectra) were not recognized by \code{QuasarNET} mainly because it fails to find at least two emission lines to be fed to its line finder units, and therefore no \code{QuasarNET} redshift was reported for them (labeled as $Z_{\code{QN}}=-1$); Fig.~\ref{fig:failedQN} represents 4 examples of these quasars. After visual inspection of these spectra, 5,190 sources were flagged as quasar and their VI redshift was reported in DR16Q catalog. \code{FNet} predicts the redshift of these 5,190 VI quasars with 87.4\% accuracy for $|\Delta\nu|< 6000~ \rm km/s$, 91.6\% accuracy for $|\Delta\nu|< 12000~ \rm km/s$ and 95.1\% accuracy for $|\Delta\nu|< 30000~ \rm km/s$; see Fig.~\ref{fig:Results} [b] and [c].  In DR16Q, \code{QuasarNET} was applied to reduce the number of quasars which need to be visually inspected, \code{FNet} has the potential to reduce more for this number. The problem of classification with \code{FNet} will be studied in the forthcoming paper.

The advantage that \code{FNet} is applicable for the spectra having unclear emission lines is due to that it finds more patterns to measure the redshift. This net finds those emission/absorption lines neglected by \code{QuasarNET} and, by purposely designed medium and large filters, it finds also the global patterns. The observed wavelength of an emission line deviates from its rest-frame wavelength because of the redshift. If the sample contains a large variation in redshift, the emission lines of a particular element will shift in a wide range of bands, then a large size mask is required to fully cover all the shifted lines.

In addition to the successful prediction of the above 5,190 quasars by \code{FNet}, these advantages can be further confirmed by testing the accuracy of \code{FNet} with a sample having no information of the prominent emission lines, including the ones considered by \code{QuasarNET}. To mask the emission lines from all the spectra, we select the size of the mask as small as possible to keep enough pixels to be fed to the network, hence we have to limit the redshift range of the samples.  In principle, this test can be carried out on wider redshift intervals. However, selecting the wider range of redshift would almost remove all pixels from the flux and no more data will be remained to be trained. Therefore, a narrow range of redshift is selected. Here, we randomly selected $11227$ quasars from DR12Q within redshift $ 2.8 \leq z \leq 3$, $90\%$ for training and $10\%$ for testing, and mask the lines including $\rm O{\rm VI}$ (103.3 nm), Ly-$\alpha$ (121.6 nm), $\rm N{\rm V}$ (124.1 nm), $\rm Si{\rm IV}+OIV$ (139.8 nm),  $\rm C{\rm IV}$ (154.9 nm) and $\rm C{\rm III}$ (190.9 nm), $\rm C{\rm II}$ (232.6 nm), an example is shown in Fig.~\ref{fig:2.8-3-mask}.

We follow the same procedure of training and testing, as described for DR16Q, for this subset of DR12Q. The velocity difference for redshift, when compared to visually inspected redshift is 98.5$\%$ accurate for $|\Delta \nu| < 6000 \rm km/s$; see Fig. \ref{fig:Results1}. This clearly declares that \code{FNet} predicts redshift, with high accuracy, even without  $\rm O{\rm VI}$, Ly-$\alpha$, $\rm N{\rm V}$, $\rm Si{\rm IV}+OIV$,  $\rm C{\rm IV}$, $\rm C{\rm III}$ and $\rm C{\rm II}$ present in the spectra. This indicates the power of the \code{FNet}, thanks to the combination of $24$ convolutional layers and the ResNet structure \citep{he2016deep}, with adopted kernel sizes of 15, 200 and 500, to identify the deeply local and global features of fluxes, respectively. Since \code{QuasarNET} requires emission lines to estimate the redshift, this test cannot be applied to it.

We also compare the \code{FNet}, spectroscopic redshift estimator, to another CNN  introduced by \citep{2018A&A...611A..97P} which estimates the photometric redshift of quasars of the Sloan Digital Sky Survey Stripe 82 \citep{2014ApJS..213...12J}. Their network takes the variability of objects into account by converting light curves into images. Their image is reproduced by assuming the five magnitudes \textit{ugriz} as the width and the height corresponding to the date of the observation of quasars. The accuracy for $|\Delta\nu|< 30000~ \rm km/s$ for their is 78.9$\%$ which for \code{FNet} for the same redshift difference is 98.9$\%$. 

\begin{figure}
\centering
\includegraphics[width=1.0\hsize,clip]{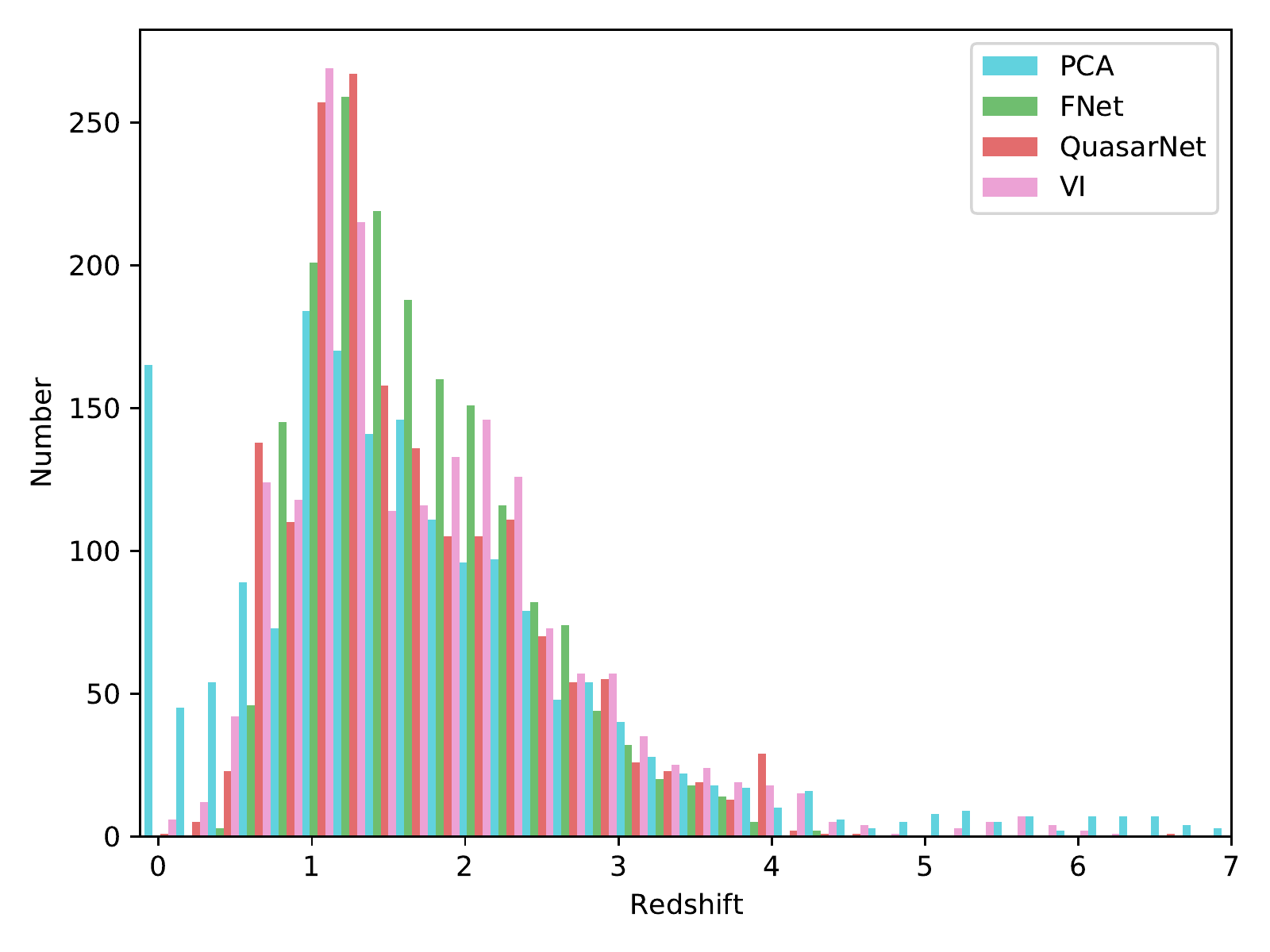}
\caption{The redshift obtained from \code{FNet} (blue), \code{QuasarNET} (orange) and PCA (red) for velocity $|\Delta\nu|> 6000~ \rm km/s$. The PCA represents some extreme estimations for redshift close to 0 or larger than 6, which are neither in agreement with the redshfits obtained from visual inspection nor from \code{FNet} and \code{QuasarNet}.}  
\label{fig:pca}
\end{figure}

Regarding the redshift obtained by the PCA method, statistically speaking, the PCA redshifts are the closest to those measured by VI and the accuracy for $|\Delta\nu|< 6000~ \rm km/s$ is 99.6$\%$. However, PCA represents some extreme estimations close to 0 or larger than 6, which are drastically different from the VI redshifts. Instead, the redshifts obtained from FNet and QuasarNet show more agreement with VI redshifts for these extreme PCA cases; see Fig~\ref{fig:pca}. Therefore, using the PCA method may cause some ambiguity in the cosmological studies of the sources in these ranges, especially the one of z>6 where the physics of the early universe becomes more important.

%

%%%%%%%%%%%%%%%%%%%%%%%%%%%%%%%%%%%%%%%%%%%%%%%%%%%%%%%%%
\section{Conclusions}\label{sec:conclusions} 
%%%%%%%%%%%%%%%%%%%%%%%%%%%%%%%%%%%%%%%%%%%%%%%%%%%%%%%%%
Deep learning, especially CNN \citep{xu2014deep,liu2018time} is promising in future Astrophysical studies. CNN can be utilized, owing to its several convolutional and fully connected layers to find the common deep hidden patterns in the spectrum (flux), in provision for improving our astrophysical knowledge of distant objects.  CNN is simple but sometimes it makes more accurate predictions than the complicated models, especially for 1--dimensional data. For instance, in the recent Kaggle competition of gravitational wave recognition \footnote{https://www.kaggle.com/c/g2net-gravitational-wave-detection/overview/description}, top three winners had adopted 1--dimensional CNN based net (ResNet \citep{he2016deep} and Inception V3 \citep{2019arXiv190503715N}). 

In this paper we represent \code{FNet}, a  CNN with ResNet structure, to estimate the redshift of quasars based on hidden patterns in the flux of quasars. \code{FNet} takes the redshift of visually inspected quasars from SDSS which have been spectrally confirmed, in the human-expert level, as quasars with confident and corrected redshift.  

The velocity difference for redshift, when compared to the visually inspected redshift is  97.0$\%$ accurate when $|\Delta \nu| < 6000 \rm km/s$, 98.0$\%$ accurate when $|\Delta \nu| < 12000 \rm km/s$, and  98.9$\%$ when $|\Delta \nu| < 30000 \rm~km/s$.

Although \code{FNet} is similar to recently adopted CNN-based redshift estimator and classifier, i.e. \code{QuasarNET} \citep{2018arXiv180809955B}, the ideas implemented in their hidden layers are different. \code{QuasarNET} follows the traditional redshift estimation procedure, namely it identifies the emission lines in spectra and based on that determines the redshift. In \code{FNet}, instead, extract the hidden patterns from the flux and consequently relates them to a specific redshift by introducing $24$ convolutional layers and the ResNet structure \citep{he2016deep} with different kernel sizes of 500, 200 and 15, without being provided any external information about emission/absorption lines. Recognizing the global pattern makes \code{FNet} outperform the \code{QuasarNET} for some complicated spectra (not enough lines, strong noise and etc.) We showed that \code{FNet} predicts the redshift of 5,190 VI quasars with 91.6\% accuracy, while \code{QuasarNET} fails to estimate. In this regard, we tested and confirmed by applying \code{FNet} on the DR12Q spectra with $\rm O{\rm VI}$ (103.3 nm), Ly-$\alpha$ (121.6 nm), $\rm N{\rm V}$ (124.1 nm), $\rm Si{\rm IV}+OIV$ (139.8 nm),  $\rm C{\rm IV}$ (154.9 nm) and $\rm C{\rm III}$ (190.9 nm), $\rm C{\rm II}$ (232.6 nm)  being removed, \code{FNet} still makes a significantly accurate redshift prediction with 99.5$\%$ for $|\Delta \nu| < 6000 \rm km/s$.

Statistically speaking \code{FNet} is capable to infer accurate redshifts even for low SNR or incomplete spectra. However, it seems that its ability to infer high redshifts is still lacking, this could be a universal disadvantage of the current deep learning methods. As it can be seen that neither \code{FNet} nor \code{QuasarNet} has inferred any quasar with redshifts greater than $6$. The main causation is the lack of training samples at the high redshift, there are only dozens of quasars at $z>6$ and less than 10 quasars at $z>7$. This problem probably could be alleviated by (a) introducing a new structure of net dedicating for small samples; (b) simulating more high redshift spectra;  (c) handling more seriously the data imbalance problem during the data preparation and the training procedure. We will investigate the prediction for high redshift sources in a separate article. Having the accurate redshift of quasars in hand, especially the high-redshift ones, will help us to modify our current knowledge of cosmology which in turn affects all areas of high energy astrophysics \citep{2012IAUS..285..312E,2020FrASS...7....8L,2019NatAs...3..272R}. 

For the classification, \code{QuasarNET} classifies the classes from the identified lines. in \code{FNet} this middle step of identifying given lines does not exist, hence it is unable to classify sources from the lines. But there is a simple way to convert our current net of inferring the redshift net to classify all SDSS spectra: changing the last output layer of \code{FNet} from one value (redshift) to several values (probabilities of all classes), and accordingly the loss function from MSE to cross-entropy. The \code{FNet} (classification) can be applied in the upcoming catalogs to reduce the number of spectra to inspect. This subject is left for the forthcoming publication.

\section*{Acknowledgment}

We thank the anonymous Referee for important remarks and comments which have improved the readability and presentation of the results. We acknowledge the support from Dr Jie Jiang and  Prof. Yifu Cai from USTC for providing the Nvidia DGX Station.

\section*{Data Availability}

The catalog underlying this article is available in The Sloan Digital Sky Survey Quasar catalog: sixteenth data release (DR16Q), at \href{https://www.sdss.org/dr16/algorithms/qso_catalog/}{https://www.sdss.org/dr16/algorithms/qsocatalog/}. The code can be found in the public domain: \href{https://github.com/AGNNet/FNet.git}{https://github.com/AGNNet/FNet.git} and the prepared data set can be found in \href{https://www.kaggle.com/ywangscience/sdss-iii-iv}{https://www.kaggle.com/ywangscience/sdss-iii-iv}.


\begin{thebibliography}{}
\makeatletter
\relax
\def\mn@urlcharsother{\let\do\@makeother \do\$\do\&\do\#\do\^\do\_\do\%\do\~}
\def\mn@doi{\begingroup\mn@urlcharsother \@ifnextchar [ {\mn@doi@}
  {\mn@doi@[]}}
\def\mn@doi@[#1]#2{\def\@tempa{#1}\ifx\@tempa\@empty \href
  {http://dx.doi.org/#2} {doi:#2}\else \href {http://dx.doi.org/#2} {#1}\fi
  \endgroup}
\def\mn@eprint#1#2{\mn@eprint@#1:#2::\@nil}
\def\mn@eprint@arXiv#1{\href {http://arxiv.org/abs/#1} {{\tt arXiv:#1}}}
\def\mn@eprint@dblp#1{\href {http://dblp.uni-trier.de/rec/bibtex/#1.xml}
  {dblp:#1}}
\def\mn@eprint@#1:#2:#3:#4\@nil{\def\@tempa {#1}\def\@tempb {#2}\def\@tempc
  {#3}\ifx \@tempc \@empty \let \@tempc \@tempb \let \@tempb \@tempa \fi \ifx
  \@tempb \@empty \def\@tempb {arXiv}\fi \@ifundefined
  {mn@eprint@\@tempb}{\@tempb:\@tempc}{\expandafter \expandafter \csname
  mn@eprint@\@tempb\endcsname \expandafter{\@tempc}}}

\bibitem[\protect\citeauthoryear{{Ajello} et~al.,}{{Ajello}
  et~al.}{2020}]{2020ApJ...892..105A}
{Ajello} M.,  et~al., 2020, \mn@doi [\apj] {10.3847/1538-4357/ab791e}, \href
  {https://ui.adsabs.harvard.edu/abs/2020ApJ...892..105A} {892, 105}

\bibitem[\protect\citeauthoryear{Albawi, Mohammed  \& Al-Zawi}{Albawi
  et~al.}{2017}]{albawi2017understanding}
Albawi S.,  Mohammed T.~A.,   Al-Zawi S.,  2017, in 2017 International
  Conference on Engineering and Technology (ICET). pp~1--6

\bibitem[\protect\citeauthoryear{Allen et~al.,}{Allen
  et~al.}{2019}]{allen2019deep}
Allen G.,  et~al., 2019, arXiv preprint arXiv:1902.00522

\bibitem[\protect\citeauthoryear{Aloysius \& Geetha}{Aloysius \&
  Geetha}{2017}]{aloysius2017review}
Aloysius N.,  Geetha M.,  2017, in 2017 International Conference on
  Communication and Signal Processing (ICCSP). pp 0588--0592

\bibitem[\protect\citeauthoryear{{Antonucci}}{{Antonucci}}{1993}]{1993ARA&A..31..473A}
{Antonucci} R.,  1993, \mn@doi [\araa] {10.1146/annurev.aa.31.090193.002353},
  \href {https://ui.adsabs.harvard.edu/abs/1993ARA&A..31..473A} {31, 473}

\bibitem[\protect\citeauthoryear{{Ba{\~n}ados} et~al.,}{{Ba{\~n}ados}
  et~al.}{2018}]{2018Natur.553..473B}
{Ba{\~n}ados} E.,  et~al., 2018, \mn@doi [\nat] {10.1038/nature25180}, \href
  {https://ui.adsabs.harvard.edu/abs/2018Natur.553..473B} {553, 473}

\bibitem[\protect\citeauthoryear{Bai, Liu, Wang  \& Yang}{Bai
  et~al.}{2018}]{bai2018machine}
Bai Y.,  Liu J.,  Wang S.,   Yang F.,  2018, The Astronomical Journal, 157, 9

\bibitem[\protect\citeauthoryear{Bailer-Jones, Irwin  \&
  Von~Hippel}{Bailer-Jones et~al.}{1998}]{bailer1998automated}
Bailer-Jones C.~A.,  Irwin M.,   Von~Hippel T.,  1998, Monthly Notices of the
  Royal Astronomical Society, 298, 361

\bibitem[\protect\citeauthoryear{Ball \& Brunner}{Ball \&
  Brunner}{2010}]{ball2010data}
Ball N.~M.,  Brunner R.~J.,  2010, International Journal of Modern Physics D,
  19, 1049

\bibitem[\protect\citeauthoryear{Bialek, Fabbro, Venn, Kumar, O'Briain  \&
  Yi}{Bialek et~al.}{2019}]{bialek2019deep}
Bialek S.,  Fabbro S.,  Venn K.~A.,  Kumar N.,  O'Briain T.,   Yi K.~M.,  2019,
  arXiv preprint arXiv:1911.02602

\bibitem[\protect\citeauthoryear{{Bolton} et~al.,}{{Bolton}
  et~al.}{2012}]{2012AJ....144..144B}
{Bolton} A.~S.,  et~al., 2012, \mn@doi [\aj] {10.1088/0004-6256/144/5/144},
  \href {https://ui.adsabs.harvard.edu/abs/2012AJ....144..144B} {144, 144}

\bibitem[\protect\citeauthoryear{{Busca} \& {Balland}}{{Busca} \&
  {Balland}}{2018}]{2018arXiv180809955B}
{Busca} N.,  {Balland} C.,  2018, arXiv e-prints, \href
  {https://ui.adsabs.harvard.edu/abs/2018arXiv180809955B} {p. arXiv:1808.09955}

\bibitem[\protect\citeauthoryear{Carleo, Cirac, Cranmer, Daudet, Schuld,
  Tishby, Vogt-Maranto  \& Zdeborov\'a}{Carleo
  et~al.}{2019}]{RevModPhys.91.045002}
Carleo G.,  Cirac I.,  Cranmer K.,  Daudet L.,  Schuld M.,  Tishby N.,
  Vogt-Maranto L.,   Zdeborov\'a L.,  2019, \mn@doi [Rev. Mod. Phys.]
  {10.1103/RevModPhys.91.045002}, 91, 045002

\bibitem[\protect\citeauthoryear{{Carroll} \& {Ostlie}}{{Carroll} \&
  {Ostlie}}{1996}]{1996ima..book.....C}
{Carroll} B.~W.,  {Ostlie} D.~A.,  1996, {An Introduction to Modern
  Astrophysics}

\bibitem[\protect\citeauthoryear{Cavuoti et~al.,}{Cavuoti
  et~al.}{2015}]{cavuoti2015machine}
Cavuoti S.,  et~al., 2015, Monthly Notices of the Royal Astronomical Society,
  452, 3100

\bibitem[\protect\citeauthoryear{{Cowie}, {Barger}, {Bauer}  \&
  {Gonz{\'a}lez-L{\'o}pez}}{{Cowie} et~al.}{2020}]{2020ApJ...891...69C}
{Cowie} L.~L.,  {Barger} A.~J.,  {Bauer} F.~E.,   {Gonz{\'a}lez-L{\'o}pez} J.,
  2020, \mn@doi [\apj] {10.3847/1538-4357/ab6aaa}, \href
  {https://ui.adsabs.harvard.edu/abs/2020ApJ...891...69C} {891, 69}

\bibitem[\protect\citeauthoryear{{Dawson} et~al.,}{{Dawson}
  et~al.}{2013}]{2013AJ....145...10D}
{Dawson} K.~S.,  et~al., 2013, \mn@doi [\aj] {10.1088/0004-6256/145/1/10},
  \href {https://ui.adsabs.harvard.edu/abs/2013AJ....145...10D} {145, 10}

\bibitem[\protect\citeauthoryear{Dawson et~al.,}{Dawson
  et~al.}{2016}]{dawson2016sdss}
Dawson K.~S.,  et~al., 2016, The Astronomical Journal, 151, 44

\bibitem[\protect\citeauthoryear{De~La~Calleja \& Fuentes}{De~La~Calleja \&
  Fuentes}{2004}]{de2004machine}
De~La~Calleja J.,  Fuentes O.,  2004, Monthly Notices of the Royal Astronomical
  Society, 349, 87

\bibitem[\protect\citeauthoryear{Dobrycheva, Vavilova, Melnyk  \&
  Elyiv}{Dobrycheva et~al.}{2017}]{dobrycheva2017machine}
Dobrycheva D.,  Vavilova I.,  Melnyk O.,   Elyiv A.,  2017, arXiv preprint
  arXiv:1712.08955

\bibitem[\protect\citeauthoryear{Duchi, Hazan  \& Singer}{Duchi
  et~al.}{2011}]{JMLR:v12:duchi11a}
Duchi J.,  Hazan E.,   Singer Y.,  2011, Journal of Machine Learning Research,
  12, 2121

\bibitem[\protect\citeauthoryear{{Ederoclite}, {Polednikova}, {Cepa}, {de Diego
  Onsurbe}  \& {Gonz{\'a}lez-Serrano}}{{Ederoclite}
  et~al.}{2012}]{2012IAUS..285..312E}
{Ederoclite} A.,  {Polednikova} J.,  {Cepa} J.,  {de Diego Onsurbe} J.~A.,
  {Gonz{\'a}lez-Serrano} I.,  2012, in {Griffin} E.,  {Hanisch} R.,   {Seaman}
  R.,  eds, New Horizons in Time Domain Astronomy. pp 312--314,
  \mn@doi{10.1017/S1743921312000907}

\bibitem[\protect\citeauthoryear{Fabbro, Venn, O'Briain, Bialek, Kielty,
  Jahandar  \& Monty}{Fabbro et~al.}{2018}]{fabbro2018application}
Fabbro S.,  Venn K.,  O'Briain T.,  Bialek S.,  Kielty C.,  Jahandar F.,
  Monty S.,  2018, Monthly Notices of the Royal Astronomical Society, 475, 2978

\bibitem[\protect\citeauthoryear{{Fan} et~al.,}{{Fan}
  et~al.}{2006}]{2006AJ....131.1203F}
{Fan} X.,  et~al., 2006, \mn@doi [\aj] {10.1086/500296}, \href
  {https://ui.adsabs.harvard.edu/abs/2006AJ....131.1203F} {131, 1203}

\bibitem[\protect\citeauthoryear{{Farr}, {Font-Ribera}  \& {Pontzen}}{{Farr}
  et~al.}{2020}]{2020JCAP...11..015F}
{Farr} J.,  {Font-Ribera} A.,   {Pontzen} A.,  2020, \mn@doi [\jcap]
  {10.1088/1475-7516/2020/11/015}, \href
  {https://ui.adsabs.harvard.edu/abs/2020JCAP...11..015F} {2020, 015}

\bibitem[\protect\citeauthoryear{Feurer \& Hutter}{Feurer \&
  Hutter}{2019}]{feurer2019hyperparameter}
Feurer M.,  Hutter F.,  2019, in , Automated Machine Learning.
Springer, Cham, pp 3--33

\bibitem[\protect\citeauthoryear{Fiorentin, Bailer-Jones, Lee, Beers, Sivarani,
  Wilhelm, Prieto  \& Norris}{Fiorentin et~al.}{2007}]{fiorentin2007estimation}
Fiorentin P.~R.,  Bailer-Jones C.,  Lee Y.~S.,  Beers T.~C.,  Sivarani T.,
  Wilhelm R.,  Prieto C.~A.,   Norris J.,  2007, Astronomy \& Astrophysics,
  467, 1373

\bibitem[\protect\citeauthoryear{{Flesch}}{{Flesch}}{2021}]{2021MNRAS.tmp..804F}
{Flesch} E.~W.,  2021, \mn@doi [\mnras] {10.1093/mnras/stab812}, \href
  {https://ui.adsabs.harvard.edu/abs/2021MNRAS.tmp..804F} {}

\bibitem[\protect\citeauthoryear{Gauci, Adami  \& Abela}{Gauci
  et~al.}{2010}]{gauci2010machine}
Gauci A.,  Adami K.~Z.,   Abela J.,  2010, arXiv preprint arXiv:1005.0390

\bibitem[\protect\citeauthoryear{Glazebrook, Offer  \& Deeley}{Glazebrook
  et~al.}{1998}]{glazebrook1998automatic}
Glazebrook K.,  Offer A.~R.,   Deeley K.,  1998, The Astrophysical Journal,
  492, 98

\bibitem[\protect\citeauthoryear{Goodfellow, Bengio  \& Courville}{Goodfellow
  et~al.}{2016}]{Goodfellow-et-al-2016}
Goodfellow I.,  Bengio Y.,   Courville A.,  2016, Deep Learning.
MIT Press

\bibitem[\protect\citeauthoryear{{Haehnelt} \& {Rees}}{{Haehnelt} \&
  {Rees}}{1993}]{1993MNRAS.263..168H}
{Haehnelt} M.~G.,  {Rees} M.~J.,  1993, \mn@doi [\mnras]
  {10.1093/mnras/263.1.168}, \href
  {https://ui.adsabs.harvard.edu/abs/1993MNRAS.263..168H} {263, 168}

\bibitem[\protect\citeauthoryear{He, Zhang, Ren  \& Sun}{He
  et~al.}{2015}]{he2015delving}
He K.,  Zhang X.,  Ren S.,   Sun J.,  2015, in Proceedings of the IEEE
  international conference on computer vision. pp 1026--1034

\bibitem[\protect\citeauthoryear{He, Zhang, Ren  \& Sun}{He
  et~al.}{2016}]{he2016deep}
He K.,  Zhang X.,  Ren S.,   Sun J.,  2016, in Proceedings of the IEEE
  conference on computer vision and pattern recognition. pp 770--778

\bibitem[\protect\citeauthoryear{Hoyle}{Hoyle}{2016}]{hoyle2016measuring}
Hoyle B.,  2016, Astronomy and Computing, 16, 34

\bibitem[\protect\citeauthoryear{{Hutchinson} et~al.,}{{Hutchinson}
  et~al.}{2016}]{2016AJ....152..205H}
{Hutchinson} T.~A.,  et~al., 2016, \mn@doi [\aj] {10.3847/0004-6256/152/6/205},
  \href {https://ui.adsabs.harvard.edu/abs/2016AJ....152..205H} {152, 205}

\bibitem[\protect\citeauthoryear{{Inayoshi}, {Visbal}  \& {Haiman}}{{Inayoshi}
  et~al.}{2020}]{2020ARA&A..58...27I}
{Inayoshi} K.,  {Visbal} E.,   {Haiman} Z.,  2020, \mn@doi [\araa]
  {10.1146/annurev-astro-120419-014455}, \href
  {https://ui.adsabs.harvard.edu/abs/2020ARA&A..58...27I} {58, 27}

\bibitem[\protect\citeauthoryear{Ioffe \& Szegedy}{Ioffe \&
  Szegedy}{2015}]{ioffe2015batch}
Ioffe S.,  Szegedy C.,  2015, in International conference on machine learning.
  pp 448--456

\bibitem[\protect\citeauthoryear{Jayalakshmi \& Santhakumaran}{Jayalakshmi \&
  Santhakumaran}{2011}]{jayalakshmi2011statistical}
Jayalakshmi T.,  Santhakumaran A.,  2011, International Journal of Computer
  Theory and Engineering, 3, 1793

\bibitem[\protect\citeauthoryear{{Jiang} et~al.,}{{Jiang}
  et~al.}{2014}]{2014ApJS..213...12J}
{Jiang} L.,  et~al., 2014, \mn@doi [\apjs] {10.1088/0067-0049/213/1/12}, \href
  {https://ui.adsabs.harvard.edu/abs/2014ApJS..213...12J} {213, 12}

\bibitem[\protect\citeauthoryear{Kim \& Brunner}{Kim \&
  Brunner}{2016}]{kim2016star}
Kim E.~J.,  Brunner R.~J.,  2016, Monthly Notices of the Royal Astronomical
  Society, p. stw2672

\bibitem[\protect\citeauthoryear{Kingma \& Ba}{Kingma \&
  Ba}{2014}]{kingma2014adam}
Kingma D.~P.,  Ba J.,  2014, arXiv preprint arXiv:1412.6980

\bibitem[\protect\citeauthoryear{Koziarski \& Cyganek}{Koziarski \&
  Cyganek}{2017}]{koziarski2017image}
Koziarski M.,  Cyganek B.,  2017, Integrated Computer-Aided Engineering, 24,
  337

\bibitem[\protect\citeauthoryear{LeCun, Bottou, Bengio  \& Haffner}{LeCun
  et~al.}{1998}]{lecun1998gradient}
LeCun Y.,  Bottou L.,  Bengio Y.,   Haffner P.,  1998, Proceedings of the IEEE,
  86, 2278

\bibitem[\protect\citeauthoryear{{Leaf} \& {Melia}}{{Leaf} \&
  {Melia}}{2019}]{2019MNRAS.487.2030L}
{Leaf} K.,  {Melia} F.,  2019, \mn@doi [\mnras] {10.1093/mnras/stz1396}, \href
  {https://ui.adsabs.harvard.edu/abs/2019MNRAS.487.2030L} {487, 2030}

\bibitem[\protect\citeauthoryear{Li, Pan  \& Duan}{Li
  et~al.}{2017}]{li2017parameterizing}
Li X.-R.,  Pan R.-Y.,   Duan F.-Q.,  2017, Research in Astronomy and
  Astrophysics, 17, 036

\bibitem[\protect\citeauthoryear{Liu, Hsaio  \& Tu}{Liu
  et~al.}{2018}]{liu2018time}
Liu C.-L.,  Hsaio W.-H.,   Tu Y.-C.,  2018, IEEE Transactions on Industrial
  Electronics, 66, 4788

\bibitem[\protect\citeauthoryear{{Lupi}, {Volonteri}, {Decarli}, {Bovino},
  {Silk}  \& {Bergeron}}{{Lupi} et~al.}{2019}]{2019MNRAS.488.4004L}
{Lupi} A.,  {Volonteri} M.,  {Decarli} R.,  {Bovino} S.,  {Silk} J.,
  {Bergeron} J.,  2019, \mn@doi [\mnras] {10.1093/mnras/stz1959}, \href
  {https://ui.adsabs.harvard.edu/abs/2019MNRAS.488.4004L} {488, 4004}

\bibitem[\protect\citeauthoryear{{Lusso}}{{Lusso}}{2020}]{2020FrASS...7....8L}
{Lusso} E.,  2020, \mn@doi [Frontiers in Astronomy and Space Sciences]
  {10.3389/fspas.2020.00008}, \href
  {https://ui.adsabs.harvard.edu/abs/2020FrASS...7....8L} {7, 8}

\bibitem[\protect\citeauthoryear{{Lyke} et~al.,}{{Lyke}
  et~al.}{2020}]{2020ApJS..250....8L}
{Lyke} B.~W.,  et~al., 2020, \mn@doi [\apjs] {10.3847/1538-4365/aba623}, \href
  {https://ui.adsabs.harvard.edu/abs/2020ApJS..250....8L} {250, 8}

\bibitem[\protect\citeauthoryear{{Madau} \& {Rees}}{{Madau} \&
  {Rees}}{2001}]{2001ApJ...551L..27M}
{Madau} P.,  {Rees} M.~J.,  2001, \mn@doi [\apjl] {10.1086/319848}, \href
  {https://ui.adsabs.harvard.edu/abs/2001ApJ...551L..27M} {551, L27}

\bibitem[\protect\citeauthoryear{{Moradi}, {Rueda}, {Ruffini}  \&
  {Wang}}{{Moradi} et~al.}{2021}]{2021A&A...649A..75M}
{Moradi} R.,  {Rueda} J.~A.,  {Ruffini} R.,   {Wang} Y.,  2021, \mn@doi [\aap]
  {10.1051/0004-6361/201937135}, \href
  {https://ui.adsabs.harvard.edu/abs/2021A&A...649A..75M} {649, A75}

\bibitem[\protect\citeauthoryear{{Mortlock} et~al.,}{{Mortlock}
  et~al.}{2011}]{2011Natur.474..616M}
{Mortlock} D.~J.,  et~al., 2011, \mn@doi [\nat] {10.1038/nature10159}, \href
  {https://ui.adsabs.harvard.edu/abs/2011Natur.474..616M} {474, 616}

\bibitem[\protect\citeauthoryear{{Ng}}{{Ng}}{2019}]{2019arXiv190503715N}
{Ng} K.,  2019, arXiv e-prints, \href
  {https://ui.adsabs.harvard.edu/abs/2019arXiv190503715N} {p. arXiv:1905.03715}

\bibitem[\protect\citeauthoryear{Odewahn, Stockwell, Pennington, Humphreys  \&
  Zumach}{Odewahn et~al.}{1992}]{odewahn1992automated}
Odewahn S.~C.,  Stockwell E.,  Pennington R.,  Humphreys R.~M.,   Zumach W.,
  1992, in , Digitised Optical Sky Surveys.
Springer, pp 215--224

\bibitem[\protect\citeauthoryear{P{\^a}ris et~al.,}{P{\^a}ris
  et~al.}{2017b}]{paris2017sloan}
P{\^a}ris I.,  et~al., 2017b, Astronomy \& Astrophysics, 597, A79

\bibitem[\protect\citeauthoryear{{P{\^a}ris} et~al.,}{{P{\^a}ris}
  et~al.}{2017a}]{2017A&A...597A..79P}
{P{\^a}ris} I.,  et~al., 2017a, \mn@doi [\aap] {10.1051/0004-6361/201527999},
  \href {https://ui.adsabs.harvard.edu/abs/2017A&A...597A..79P} {597, A79}

\bibitem[\protect\citeauthoryear{{P{\^a}ris} et~al.,}{{P{\^a}ris}
  et~al.}{2018}]{2018A&A...613A..51P}
{P{\^a}ris} I.,  et~al., 2018, \mn@doi [\aap] {10.1051/0004-6361/201732445},
  \href {https://ui.adsabs.harvard.edu/abs/2018A&A...613A..51P} {613, A51}

\bibitem[\protect\citeauthoryear{Pascanu, Mikolov  \& Bengio}{Pascanu
  et~al.}{2013}]{pascanu2013difficulty}
Pascanu R.,  Mikolov T.,   Bengio Y.,  2013, in International conference on
  machine learning. pp 1310--1318

\bibitem[\protect\citeauthoryear{{Pasquet-Itam} \& {Pasquet}}{{Pasquet-Itam} \&
  {Pasquet}}{2018}]{2018A&A...611A..97P}
{Pasquet-Itam} J.,  {Pasquet} J.,  2018, \mn@doi [\aap]
  {10.1051/0004-6361/201731106}, \href
  {https://ui.adsabs.harvard.edu/abs/2018A&A...611A..97P} {611, A97}

\bibitem[\protect\citeauthoryear{{Paszke} et~al.,}{{Paszke}
  et~al.}{2019}]{2019arXiv191201703P}
{Paszke} A.,  et~al., 2019, arXiv e-prints, \href
  {https://ui.adsabs.harvard.edu/abs/2019arXiv191201703P} {p. arXiv:1912.01703}

\bibitem[\protect\citeauthoryear{{Pedregosa} et~al.,}{{Pedregosa}
  et~al.}{2012}]{2012arXiv1201.0490P}
{Pedregosa} F.,  et~al., 2012, arXiv e-prints, \href
  {https://ui.adsabs.harvard.edu/abs/2012arXiv1201.0490P} {p. arXiv:1201.0490}

\bibitem[\protect\citeauthoryear{{P{\'e}rez-R{\`a}fols} \&
  {Pieri}}{{P{\'e}rez-R{\`a}fols} \& {Pieri}}{2020}]{2020MNRAS.496.4941P}
{P{\'e}rez-R{\`a}fols} I.,  {Pieri} M.~M.,  2020, \mn@doi [\mnras]
  {10.1093/mnras/staa1786}, \href
  {https://ui.adsabs.harvard.edu/abs/2020MNRAS.496.4941P} {496, 4941}

\bibitem[\protect\citeauthoryear{{P{\'e}rez-R{\`a}fols}, {Pieri}, {Blomqvist},
  {Morrison}  \& {Som}}{{P{\'e}rez-R{\`a}fols}
  et~al.}{2020}]{2020MNRAS.496.4931P}
{P{\'e}rez-R{\`a}fols} I.,  {Pieri} M.~M.,  {Blomqvist} M.,  {Morrison} S.,
  {Som} D.,  2020, \mn@doi [\mnras] {10.1093/mnras/stz3467}, \href
  {https://ui.adsabs.harvard.edu/abs/2020MNRAS.496.4931P} {496, 4931}

\bibitem[\protect\citeauthoryear{Refaeilzadeh, Tang  \& Liu}{Refaeilzadeh
  et~al.}{2009}]{refaeilzadeh2009cross}
Refaeilzadeh P.,  Tang L.,   Liu H.,  2009, Encyclopedia of database systems,
  5, 532

\bibitem[\protect\citeauthoryear{{Ricci} et~al.,}{{Ricci}
  et~al.}{2017}]{2017ApJS..233...17R}
{Ricci} C.,  et~al., 2017, \mn@doi [\apjs] {10.3847/1538-4365/aa96ad}, \href
  {https://ui.adsabs.harvard.edu/abs/2017ApJS..233...17R} {233, 17}

\bibitem[\protect\citeauthoryear{{Risaliti} \& {Lusso}}{{Risaliti} \&
  {Lusso}}{2019}]{2019NatAs...3..272R}
{Risaliti} G.,  {Lusso} E.,  2019, \mn@doi [Nature Astronomy]
  {10.1038/s41550-018-0657-z}, \href
  {https://ui.adsabs.harvard.edu/abs/2019NatAs...3..272R} {3, 272}

\bibitem[\protect\citeauthoryear{Sadeh, Abdalla  \& Lahav}{Sadeh
  et~al.}{2016}]{sadeh2016annz2}
Sadeh I.,  Abdalla F.~B.,   Lahav O.,  2016, Publications of the Astronomical
  Society of the Pacific, 128, 104502

\bibitem[\protect\citeauthoryear{{Schneider} et~al.,}{{Schneider}
  et~al.}{2002}]{2002AJ....123..567S}
{Schneider} D.~P.,  et~al., 2002, \mn@doi [\aj] {10.1086/338434}, \href
  {https://ui.adsabs.harvard.edu/abs/2002AJ....123..567S} {123, 567}

\bibitem[\protect\citeauthoryear{{Schneider} et~al.,}{{Schneider}
  et~al.}{2010}]{2010AJ....139.2360S}
{Schneider} D.~P.,  et~al., 2010, \mn@doi [\aj] {10.1088/0004-6256/139/6/2360},
  \href {https://ui.adsabs.harvard.edu/abs/2010AJ....139.2360S} {139, 2360}

\bibitem[\protect\citeauthoryear{Sengupta, Basak, Saikia, Paul, Tsalavoutis,
  Atiah, Ravi  \& Peters}{Sengupta et~al.}{2020}]{sengupta2020review}
Sengupta S.,  Basak S.,  Saikia P.,  Paul S.,  Tsalavoutis V.,  Atiah F.,  Ravi
  V.,   Peters A.,  2020, Knowledge-Based Systems, 194, 105596

\bibitem[\protect\citeauthoryear{Sharma, Kembhavi, Kembhavi, Sivarani, Abraham
  \& Vaghmare}{Sharma et~al.}{2020}]{sharma2020application}
Sharma K.,  Kembhavi A.,  Kembhavi A.,  Sivarani T.,  Abraham S.,   Vaghmare
  K.,  2020, Monthly Notices of the Royal Astronomical Society, 491, 2280

\bibitem[\protect\citeauthoryear{Tieleman \& Hinton}{Tieleman \&
  Hinton}{2012}]{Tieleman}
Tieleman T.,  Hinton 2012, Lecture 6.5 - RMSProp, COURSERA

\bibitem[\protect\citeauthoryear{Tietz, Fan, Nouri, Bossan  \& {skorch
  Developers}}{Tietz et~al.}{2017}]{skorch}
Tietz M.,  Fan T.~J.,  Nouri D.,  Bossan B.,   {skorch Developers} 2017,
  skorch: A scikit-learn compatible neural network library that wraps PyTorch.
\url {https://skorch.readthedocs.io/en/stable/}

\bibitem[\protect\citeauthoryear{{Wang} et~al.,}{{Wang}
  et~al.}{2021}]{2021ApJ...907L...1W}
{Wang} F.,  et~al., 2021, \mn@doi [\apjl] {10.3847/2041-8213/abd8c6}, \href
  {https://ui.adsabs.harvard.edu/abs/2021ApJ...907L...1W} {907, L1}

\bibitem[\protect\citeauthoryear{{Willott} et~al.,}{{Willott}
  et~al.}{2010}]{2010AJ....140..546W}
{Willott} C.~J.,  et~al., 2010, \mn@doi [\aj] {10.1088/0004-6256/140/2/546},
  \href {https://ui.adsabs.harvard.edu/abs/2010AJ....140..546W} {140, 546}

\bibitem[\protect\citeauthoryear{Xu, Ren, Liu  \& Jia}{Xu
  et~al.}{2014}]{xu2014deep}
Xu L.,  Ren J.~S.,  Liu C.,   Jia J.,  2014, Advances in neural information
  processing systems, 27, 1790

\bibitem[\protect\citeauthoryear{Xu, Wang, Chen  \& Li}{Xu
  et~al.}{2015}]{xu2015empirical}
Xu B.,  Wang N.,  Chen T.,   Li M.,  2015, arXiv preprint arXiv:1505.00853

\bibitem[\protect\citeauthoryear{Yamashita, Nishio, Do  \& Togashi}{Yamashita
  et~al.}{2018}]{yamashita2018convolutional}
Yamashita R.,  Nishio M.,  Do R. K.~G.,   Togashi K.,  2018, Insights into
  imaging, 9, 611

\bibitem[\protect\citeauthoryear{Yang, Nguyen, San, Li  \& Krishnaswamy}{Yang
  et~al.}{2015}]{yang2015deep}
Yang J.,  Nguyen M.~N.,  San P.~P.,  Li X.,   Krishnaswamy S.,  2015, in Ijcai.
  pp 3995--4001

\bibitem[\protect\citeauthoryear{{Yang} et~al.,}{{Yang}
  et~al.}{2020}]{2020ApJ...897L..14Y}
{Yang} J.,  et~al., 2020, \mn@doi [\apjl] {10.3847/2041-8213/ab9c26}, \href
  {https://ui.adsabs.harvard.edu/abs/2020ApJ...897L..14Y} {897, L14}

\makeatother
\end{thebibliography}
\end{document}